\documentclass[preprint2]{aastex63}
\usepackage[whole]{bxcjkjatype} 
\usepackage[utf8]{inputenc}
\accepted{\today}

\shorttitle{Particle acceleration in BH magnetospheres}
\shortauthors{Hirotani et al.}
\graphicspath{{./}{figures/}}

\begin{document}

\title{Two-dimensional Particle-in-Cell simulations of axisymmetric black hole magnetospheres}


\correspondingauthor{Kouichi Hirotani}
\email{hirotani@asiaa.sinica.edu}

\author[0000-0002-2472-9002]{Kouichi Hirotani}
\affiliation{Institute of Astronomy and Astrophysics, Academia Sinica, 
Taipei 10617, Taiwan, R.O.C.}

\author[0000-0001-5557-5387]{Ruben Krasnopolsky}
\affiliation{Institute of Astronomy and Astrophysics, Academia Sinica, 
Taipei 10617, Taiwan, R.O.C.}

\author[0000-0001-8385-9838]{Hsien Shang （尚賢）}
\affiliation{Institute of Astronomy and Astrophysics, Academia Sinica, 
Taipei 10617, Taiwan, R.O.C.}

\author[0000-0001-6031-7040]{Ken-ichi Nishikawa}
\affiliation{Department of Physics, Chemistry and Mathematics, 
Alabama A\&M University, Huntsville, AL 35811, USA
}


\author[0000-0001-5783-0513]{Michael Watson}
\affiliation{Department of Physics, Fisk University, 1000 17th Ave North, 
Nashville, TN 37208, USA}




\begin{abstract}
We investigate the temporal evolution of an axisymmetric magnetosphere
around a rapidly rotating, stellar-mass black hole,
applying a two-dimensional particle-in-cell simulation scheme.
Adopting a homogeneous pair production,
and assuming that the mass accretion rate 
is much less than the Eddington limit,
we find that the black hole's rotational energy
is preferentially extracted from the middle latitudes,
and that this outward energy flux exhibits 
an enhancement that lasts approximately 160 dynamical time scales.
It is demonstrated that the magnetohydrodynamic approximations
cannot be justified in such a magnetically-dominated magnetosphere, 
because the Ohm's law completely breaks down, and
because the charge-separated electron-positron plasmas are 
highly non-neutral.
An implication is given regarding the collimation of
relativistic jets.
\end{abstract}



\section{Introduction}
\label{sec:intro}
It is commonly accepted that every active galaxy harbors a
super-massive black hole (BH) in its center
\citep{ferra96,larkin16,kormendy13}.
A likely mechanism for powering their relativistic jets 
is the release of the rotational energy of the BHs
\citep{bla77},
which is referred to as the Blandford-Znajek (BZ) process,
as demonstrated by general-relativistic (GR) magnetohydrodynamic 
simulations 
\citep{Koide:2002:Sci,McKinney:2012:MNRAS}.
In the polar region of a BH magnetosphere,
centrifugal force prevents accretion toward the rotation axis 
and a high vacuum is maintained
\citep{hirose04}.
In this nearly vacuum BH magnetosphere, electrons and positrons ($e^{\pm}$'s) 
are supplied via the collisions of MeV photons emitted from the equatorial, 
advection-dominated accretion flow (ADAF)
\citep{ichimaru77,narayan94}.
Particularly, when the mass accretion rate is much less than
the Eddington rate, the collisions can no longer sustain 
a force-free magnetosphere
\citep{levi11}.
In such a charge-starved magnetosphere, an electric field appears 
along the magnetic field lines. 
In this vacuum gap, a portion of the BZ flux is dissipated 
as particle acceleration and radiation
\citep{bes92,Hirotani:1998:ApJ,nero07,hiro16a}. 
Such a highly vacuum BH magnetosphere 
has been investigated by the particle-in-cell (PIC) 
scheme one-dimensionally along a radial magnetic field line
\citep{Levinson:2018:AA,Chen:2018:ApJL,Chen:2020:ApJ,Kisaka:2020:arXiv},
and two-dimensionally in the poloidal plane
\citep{Parfrey:2019:PhRvL,Crinquand:2020:PhRvL}.
In the present paper, 
adopting a fixed magnetic field in the poloidal plane,
we perform a two-dimensional, axisymmetric PIC simulation
around a stellar-mass BH.
We will examine the temporal evolution 
of the poloidal components of the electric field,
toroidal component of the magnetic field, 
and the distribution functions of the 
electrons and positrons that are created
and accelerated within the BH mangetosphere.

After describing the background spacetime in \S~\ref{sec:geometry}, 
we consider the initial conditions of the PIC simulation
in \S~\ref{sec:stationary}.
Then we present the results of our 2-D GR PIC simulation 
in \S~\ref{sec:PIC},
demonstrating that the magnetohydrodynamic (MHD)
approximations totally break down 
in the vicinity (i.e., at the jet-launching region)
of rotating BHs that accrete plasmas at much lower rate than the Eddington rate.
In the present paper we focus on a set of astrophysical cases, as described in \S~\ref{sec:nonstationary}.
Code testing cases, including comparison with exact solutions (e.g., development of the electromagnetic field in a plasma-free spacetime, or propagations of charged particles in a fixed electromagnetic field) have been performed, but will be deferred to subsequent papers. 
Finally, in \S~\ref{sec:disc}
we discuss the implication of astrophysical applications to 
the very-long-baseline-interferometric (VLBI) 
observations of supermassive BHs in the center of
low-luminosity active galactic neclei.

\section{Stationary magnetosphere} 
\label{sec:stationary}
Let us begin with the stationary solution of the electromagnetic fields
in a magnetosphere of a rotating black hole.
We will use this stationary solution as the initial condition
of time-dependent particle-in-cell (PIC) simulations.
Throughout the present paper, we consider two-dimentional structure
of the BH magnetosphere in which electrons and positrons are
created and accelerated,
assuming axisymmetry with respect to the rotation axis of the BH.


\subsection[]{Background geometry}
\label{sec:geometry}
Around a rotating, non-charged BH, 
the background geometry is described by the Kerr metric
\citep{kerr63}.
In the Boyer-Lindquist coordinates \citep{boyer67}, 
the line element can be expressed as
\begin{equation}
 ds^2= g_{tt} dt^2
      +2g_{t\varphi} dt d\varphi
      +g_{\varphi\varphi} d\varphi^2
      +g_{rr} dr^2
      +g_{\theta\theta} d\theta^2,
  \label{eq:metric}
\end{equation}
where 
\begin{equation}
   g_{tt} 
   \equiv 
   -\frac{\Delta-a^2\sin^2\theta}{\Sigma},
   \qquad
   g_{t\varphi}
   \equiv 
   -\frac{2Mar \sin^2\theta}{\Sigma}, 
  \label{eq:metric_2}
\end{equation}
\begin{equation}
   g_{\varphi\varphi}
     \equiv 
     \frac{A \sin^2\theta}{\Sigma} , 
     \qquad
   g_{rr}
     \equiv 
     \frac{\Sigma}{\Delta} , 
     \qquad
   g_{\theta\theta}
     \equiv 
     \Sigma ;
  \label{eq:metric_3}
\end{equation}
$\Delta \equiv r^2-2Mr+a^2$,
$\Sigma\equiv r^2 +a^2\cos^2\theta$,
$A \equiv (r^2+a^2)^2-\Delta a^2\sin^2\theta$.
In equations~(\ref{eq:metric})--(\ref{eq:metric_3}), 
we adopt the geometrized unit, putting $c=G=1$, where
$c$ and $G$ denote the speed of light and the gravitational constant,
The horizon radius, $r_{\rm H} \equiv M+\sqrt{M^2-a^2}$,
is obtained by $\Delta=0$, 
where $M$ corresponds to the gravitational radius.
The spin parameter becomes $a=M$ for a maximally rotating BH,
and $a=0$ for a non-rotating BH. 

\subsection[]{The Zero Angular Momentum Observer (ZAMO)}
\label{sec:ZAMO}
As a fiducial observer,
let us introduce the Zero Angular Momentum Observer (ZAMO),
who is static in the poloidal plane ($r$,$\theta$)
but rotates around the BH at the same angular frequency
as the spacetime frame dragging frequency,
$\omega \equiv -g_{t\varphi}/g_{\varphi\varphi}$.
The tetrad of ZAMO reads
\begin{equation}
  \tilde{\mbox{\boldmath$e$}}_{(\hat{t})}
  = \alpha^{-1} 
    (\mbox{\boldmath$e$}_{(t)}
     +\omega \mbox{\boldmath$e$}_{(\varphi)}),
  \label{eq:rotationg_e0z}
\end{equation}
\begin{equation}
  \tilde{\mbox{\boldmath$e$}}_{(\hat{r})}
  = \sqrt{\frac{\Delta}{\Sigma}} 
    \mbox{\boldmath$e$}_{(r)},
  \label{eq:rotationg_e1z}
\end{equation}
\begin{equation}
  \tilde{\mbox{\boldmath$e$}}_{(\hat{\theta})}
  = \frac{1}{\sqrt{\Sigma}} 
    \mbox{\boldmath$e$}_{(\theta)},
  \label{eq:rotationg_e2z}
\end{equation}
\begin{equation}
  \tilde{\mbox{\boldmath$e$}}_{(\hat{\varphi})}
  = \frac{1}{\sqrt{g_{\varphi\varphi}}}
    \mbox{\boldmath$e$}_{(\varphi)},
  \label{eq:rotationg_e3z}
\end{equation}
where 
\begin{equation}
  \alpha 
  \equiv \frac{d\tau}{dt}
  = \frac{\rho_{\rm w}}{\sqrt{g_{\varphi\varphi}}}
  = \sqrt{\frac{\Delta\Sigma}{A}}
  \label{eq:lapse}
\end{equation}
denotes the lapse; 
$d\tau$ refers to the ZAMO proper time.
The tilde (\,$\tilde{}$\,) represents a ZAMO-measured quantity
and the caret (\,$\hat{}$\,) does that the component is projected
on an orthnormal basis.
At the horizon we have $\alpha=0$, 
while away from the BH we have $\alpha=1$.
The coordinate bases are defined as
\begin{equation}
  \mbox{\boldmath$e$}_{(t)} = \partial_t,
  \quad 
  \mbox{\boldmath$e$}_{(r)} = \partial_r,
  \quad 
  \mbox{\boldmath$e$}_{(\theta)} = \partial_\theta,
  \quad 
  \mbox{\boldmath$e$}_{(\varphi)} = \partial_\varphi.
  \label{eq:coord_bases}
\end{equation}

%

We use the ZAMO to solve the electromagnetic fields in a stationary 
magnetosphere (\S~\ref{sec:stationary_laws}), 
as well as to present the current densities
(figs.~\ref{fig:J1J2} \& \ref{fig:chJp})
and the particle distribution functions 
(figs.~\ref{fig:Distr_LF} \& \ref{fig:Distr_PA})
in a non-stationary magnetosphere (\S~\ref{sec:PIC}).

\subsection{Gauss's and Biot-Savart laws}
\label{sec:stationary_laws}
To describe the stationary electromagnetic field,
we should simultaneously solve
the Gauss's law and the Biot \& Savart law.
The expressions of these two laws are derived
in the Appendix~\ref{sec:appendix_stationary}.
For convenience, we replace the independent variable $r$
with the so-called \lq\lq tortoise coordinate'', $r_\ast$.
It is defined by
\begin{equation}
  \frac{d r_\ast}{dr}=\frac{r^2+a^2}{\Delta}.
  \label{eq:tortoise}
\end{equation}
In this coordinate, the horizon corresponds to
$r_\ast \rightarrow -\infty$.
At large distances, $dr_\ast/dr \rightarrow 1$.

What is more, to avoid the sigular behaviour due to the
$\csc\theta$ factors in 
the Gauss's law (\ref{eq:Poisson_3}) and 
the Biot and Savart law (\ref{eq:BS_2})
at the poles (i.e., at $\theta=0$ and $\pi$),
we introduce a new meridional variable, $y \equiv 1-\cos\theta$.
Adopting this $y$ coordinate, we obtain
\begin{equation}
  \frac{1}{\sin\theta}\frac{\partial}{\partial\theta}
  = \frac{\partial}{\partial y}
  \label{eq:xi}
\end{equation}

To avoid the change of the type of the second-order
partial differential equation 
(specifically, the Biot and Savart law) 
at the static limit,
we replace the electro-static potential $A_t$
with the ZAMO-measured value, $A_{\hat{t}}$.
The tetrad of ZAMO
(eqs.~[\ref{eq:rotationg_e0z}]--[\ref{eq:rotationg_e3z}]) 
gives
\begin{equation}
  A_t= \alpha A_{\hat{t}} - \omega A_\varphi
  \label{eq:At}
\end{equation}
and
\begin{equation}
  A_\varphi= \sqrt{g_{\varphi\varphi}} A_{\hat{\varphi}}.
  \label{eq:Aphi}
\end{equation}
Using $A_{\hat{t}}$, $A_\varphi$, $r_\ast$ and $y$,
the Biot and Savart law (eq.~[\ref{eq:BS_2}]) can be rewritten as
\begin{eqnarray}
 -\lefteqn{
  \frac{\Delta\Sigma}{A}
  \frac{\partial^2 A_\varphi}{\partial r_\ast{}^2}
  + C_1 
  \frac{\partial A_\varphi}{\partial r_\ast}
 -\frac{\Delta^2 \Sigma \sin^2\theta}
       {(r^2+a^2)^2 A}
  \frac{\partial^2 A_\varphi}{\partial y^2}
          }
  \nonumber\\ 
 &&
 +\left( \frac{\Delta}{r^2+a^2} \right)^2 
  \frac{4 M a^2 r \sin^2\theta \cos\theta}
       {\Sigma A}
  \frac{\partial A_\varphi}{\partial y}
 +C_0 A_\varphi
  \nonumber\\ 
 &&
 +\frac{2 M a r \sin^2\theta}{\Sigma}
  \frac{\partial^2}{\partial r_\ast{}^2} (\alpha A_{\hat{t}})
 + C_2
  \frac{\partial}{\partial r_\ast} (\alpha A_{\hat{t}})
  \nonumber\\ 
 &&
 +\frac{2 M a r \Delta \sin^2\theta}
       {(r^2+a^2)^2 \Sigma}
  \frac{\partial^2}{\partial y^2}(\alpha A_{\hat{t}})
  \nonumber\\
&&
 +\frac{4 M a r \Delta \sin^2\theta \cos\theta}
       {(r^2+a^2) \Sigma^2}
  \frac{\partial}{\partial y}(\alpha A_{\hat{t}})
  \nonumber\\ 
 &&
 = 4 \pi \left( \frac{\Delta}{r^2+a^2} \right)^2
   \Sigma \sin^2\theta \cdot J^\varphi,
  \label{eq:BS_3}
\end{eqnarray}
where
\begin{eqnarray}
  \lefteqn{
  C_0 
  \equiv 
  -\frac{2 M a r \Delta \sin^2\theta}{(r^2+a^2)^2 \Sigma}
  \left\{ \right.
          \Delta (\partial_r{}^2 \omega)
          }
  \nonumber\\
    && +\left[ \frac{(r^2-a^2\cos^2\theta)\Delta}{r\Sigma}
              -\frac{2M(r^2-a^2)}{r^2+a^2}
        \right]
        (\partial_r \omega) 
  \nonumber\\
    && +\sin^2\theta (\partial_y{}^2 \omega)
  \nonumber\\
    && +\frac{2(r^2+a^2)\cos\theta}{\Sigma} (\partial_y \omega)
        \left. \right\}, 
  \label{eq:def_c0}
\end{eqnarray}
\begin{eqnarray}
  \lefteqn{
  C_1 
  \equiv 
  \frac{2 \Delta}{(r^2+a^2)^2 \Sigma^2}
  }
  \nonumber\\
  &&
  \times
  \left[ \right.
        -(r-M)(r^2+a^2)\Sigma + \frac{(r^2+a^2)\Sigma}{A} C_3
  \qquad\qquad
  \nonumber\\
    && \qquad
       +(\Delta-a^2\sin^2\theta) r a^2 \sin^2\theta
       +\frac{4M^2 a^4 r^3 \sin^4\theta}{A}
  \nonumber\\
    && \qquad
       + 2 M a r \sin^2\theta (r^2+a^2) \Sigma (\partial_r \omega)
  \left. \right] , 
  \label{eq:def_c1}
\end{eqnarray}
\begin{eqnarray}
  \lefteqn{
    C_2 
    \equiv 
    \frac{2 M a \sin^2\theta}{(r^2+a^2)^2 \Sigma^2}
    } \nonumber\\
  && \times
  \left[ \right.
        -r^6 +a^2(-2+\cos^2\theta) r^4
        + 4Ma^2 \sin^2\theta r^3
  \nonumber\\
  &&  \qquad  +a^4 (1-2\sin^2\theta) r^2 + a^6 \cos^2\theta
  \left. \right] ,
  \label{eq:def_c2}
\end{eqnarray}
\begin{eqnarray}
  \lefteqn{
  C_3 \equiv  (r-M)A +a^2 \sin^2\theta 
          }
  \nonumber\\ 
  && \times
  \left[ -r^3 -3Mr^2+(2M^2-a^2\cos^2\theta)r
         +a^2 M \cos^2\theta \right] .
  \nonumber\\
  \label{eq:def_c3}
\end{eqnarray}
It follows that all the four highest-order derivative terms
of equation~(\ref{eq:BS_3}) have definite signs;
thus, the Biot-Savart law
now become an elliptic type 
in the entire simulation region,
because we adopt a physical observer, ZAMO.

The Gauss's law~(\ref{eq:Poisson_3}) can also be re-written
with respect to $\alpha A_{\hat{t}}$ and $A_\varphi$.
Multiplying $\left[ \Delta / (r^2+a^2) \right]^2$ on both sides,
we obtain
\begin{eqnarray}
 &&
  \frac{A}{\Sigma^2} 
  \frac{\partial^2 (\alpha A_{\hat{t}})}{\partial r_\ast^2}
 +D_1
  \frac{\partial (\alpha A_{\hat{t}})}{\partial r_\ast}
 -\frac{\Delta A}{(r^2+a^2)\Sigma^2} 
  \left( \partial_r \omega \right) 
  \frac{\partial A_\varphi}{\partial r_\ast}
 \nonumber\\
 &&
 +\frac{\Delta A \sin^2\theta}{(r^2+a^2)^2 \Sigma^2}
  \frac{\partial^2 (\alpha A_{\hat{t}})}{\partial \xi^2}
 \nonumber\\
 &&
 +\frac{2\Delta \cos\theta}{(r^2+a^2)^2 \Sigma^3}
  \left[ (r^2+a^2)A-a^2 \Delta \Sigma \sin^2\theta \right]
  \frac{\partial (\alpha A_{\hat{t}})}{\partial \xi}
 \nonumber\\
 &&
 +\frac{2 \Delta \sin^2\theta}{(r^2+a^2)^2 \Sigma^2}
  \left[ -A (\partial_\xi \omega) +\Delta a^2\cos\theta \cdot \omega
  \right]
  \frac{\partial A_\varphi}{\partial \xi}
  \nonumber\\
 &&
 + D_0 A_\varphi
 = 4\pi \rho \left( \frac{\Delta}{r^2+a^2} \right)^2 ,
 \label{eq:Poisson_4}
\end{eqnarray}
where
\begin{eqnarray}
 \lefteqn{
 D_0
 \equiv
  \frac{\Delta A}{(r^2+a^2)^2 \Sigma^2} 
  \biggl\{ \biggr.
    -\Delta (\partial_r{}^2 \omega)
    -\sin^2\theta (\partial_\xi{}^2 \omega)
         }
  \nonumber\\
  &&
    +\left[ 2(r-M)-\Delta \partial_r \left( \ln\frac{\Delta A}
                                                   {\Sigma}
                                     \right)
     \right]
     (\partial_r \omega)
  \nonumber\\
  &&
  -\frac{2\cos\theta}{\Sigma}
   \left[ r^2+a^2-\frac{\Delta \Sigma}{A} a^2 \sin^2\theta
   \right]
   (\partial_\xi \omega)
  \biggl.
  \biggr\} \quad
  \label{eq:Def_D0}
\end{eqnarray}
and
\begin{eqnarray}
 \lefteqn{ 
  D_1
  \equiv
 -\frac{A}{(r^2+a^2)\Sigma^2} 
         }
  \nonumber\\
  && \times
  \left[ 2(r-M)-\Delta \partial_r 
                \left( \ln \frac{(r^2+a^2)A}{\Sigma}
                \right)
  \right]
  \label{eq:Def_D1}
\end{eqnarray}

Note that $\partial^2 A_\varphi / \partial r_\ast{}^2$ and
$\partial^2 A_\varphi / \partial \xi^2$ terms vanish
in equation~(\ref{eq:Poisson_4}).

\subsection[]{Boundary conditions}
\label{sec:BDCs}
We search for the stationary solution 
that satisfy the Gauss's law (\ref{eq:Poisson_4}) and 
the Biot and Savart law (\ref{eq:BS_3}).
To this end,
we must impose boundary conditions 
$\alpha A_{\hat{t}}$ and $A_\varphi$,
or equivalently on $A_t$ and $A_\varphi$.
In the present paper, 
we solve these two equations in the first and fourth quadrants of the poloidal plane ($r$,$\theta$).
The region is bordered by \\
$\bullet$
the polar boundary at $\theta=0$ (north polar axis) and 
at $\theta=\pi$ (south polar axis),\\
$\bullet$
the inner boundary at $r=r_{\rm H}$, and \\
$\bullet$
the outer oundary at $r=r_{\rm out}$, 
where $r_{\rm out} \gg r_{\rm g} \equiv GMc^{-2}=M$.\\
In this subsection, we describe the boundary conditions 
at these three boundaries.

At the northern and southern polar boundaries, 
we impose
$E_\theta=\partial_\theta A_t=0$, that is, a Neumann condition on $A_t$.
We also impose
$B^\theta \propto F_{\varphi r}=-\partial_r A_\varphi=0$.
Thus, we put $A_\varphi=0$ at $\theta=0$ and $\pi$ 
and measure the magnetic flux function $A_\varphi$ 
from the rotation axis.

At the inner boundary, we impose 
$\mbox{\boldmath$E$} \cdot \mbox{\boldmath$B$} =0$
and $F_{\varphi r}=0$,
where $F^{r\theta}=0$ is assumed at $t=0$.
For instance, in ZAMO, these conditions constrain
that both the radial component of the electric field
and the meridional component of the magnetic field
vanish at the inner boundary.

At the outer boundary
($r=r_{\rm out} \gg M$), 
we impose
$\mbox{\boldmath$E$} \cdot \mbox{\boldmath$B$} 
    \propto (\partial_r A_t) (\partial_\theta A_\varphi)
           -(\partial_\theta A_t) (\partial_r A_\varphi)
    =0$
and $\partial_r A_\varphi=0$,
the latter of which comes from the assumption of 
a split-monopole magnetic field,
$J^\varphi{}_{\rm eq} \propto r^{-4}$.
Thus, in the present paper we impose 
the Neumann condition, $\partial_r A_t=0$.
However, in general, if we impose
the magnetic field direction, 
$\partial_r A_\varphi / \partial_\theta A_\varphi$
(e.g., if we adopt a paraboloidal magnetic field,
see BZ77 for details),
$\mbox{\boldmath$E$} \cdot \mbox{\boldmath$B$} =0$
constrains the direction of the 
$A_t={\rm constant}$ surface at the outer boundary.

\subsection[]{Disk toroidal current}
\label{sec:disk_current}
We assume that the plasmas in an ADAF produce
a toroidal current 
$J^\varphi{}_{\rm eq} = C_{\rm eq} r^{-4}$ 
near the equator 
all the way to the horizon
within the colatitudes 
$87^\circ < \theta < 93^\circ$;
outside of this equatorial region, 
$J^\varphi{}_{\rm eq}=0$ is assumed.
For a slowly rotating BH, this disk current produces
a split-monopole magnetic field (BZ77).

The normalization factor $C_{\rm eq}$ is adjusted
so that the meridionally averaged, 
ZAMO-measured radial magnetic field strength 
at $r=2M$,
\begin{equation}
  \langle B^r(2M) \rangle 
    \equiv 
    \frac{\int_0^\pi \tilde{B}^{\hat{r}}(2M,\theta) \sqrt{A}\sin\theta d\theta}
                {\int_0^\pi               \sqrt{A}\sin\theta d\theta},
  \label{eq:Br_avr}
\end{equation}
may match a fraction of the equipartition field strength
\citep{Yuan:2014:ARA&A},
\begin{equation}
  B_{\rm eq}(r)
  = 9.7 \times 10^7 
    \left( \frac{\dot{m}}{M_1} \right)^{1/2} 
    \left( \frac{r}{2M} \right)^{-5/4} 
    \mbox{ G}, 
  \label{eq:B_eq}
\end{equation}
which is obtained if there is an equipartition between
the magnetic field energy density and the plasma internal energy density;
the alpha viscous parameter is assumed to be $0.3$.
The dimensionless accretion rate 
$\dot{m}$ is defined by
\begin{equation}
  \dot{m} \equiv \frac{\dot{M}}{\dot{M}_{\rm Edd}},
\end{equation}
where $\dot{M}$ denotes the mass accretion rate.
The Eddington accretion rate is defined by
\begin{equation}
  \dot{M}_{\rm Edd} 
  \equiv 
  \frac{L_{\rm Edd}}{\eta_{\rm eff} c^2}
  = 1.39 \times 10^{19} M_1 {\rm g \ s}^{-1},
\end{equation}
where $L_{\rm Edd}$ denotes 
the Eddington luminosity;
the conversion efficiency is assumed to be 
$\eta_{\rm eff}=0.1$.

In the present paper,
we adopt a relatively small mass accretion rate, 
$\dot{m}=2.25 \times 10^{-4}$,
and $\langle B^r(2M) \rangle = B_{\rm eq}(2M)$.

\subsection[]{Stationary solution}
\label{sec:stationary_EM}
\begin{figure*}
\hspace{4.3cm}
\includegraphics[width=0.5\textwidth]{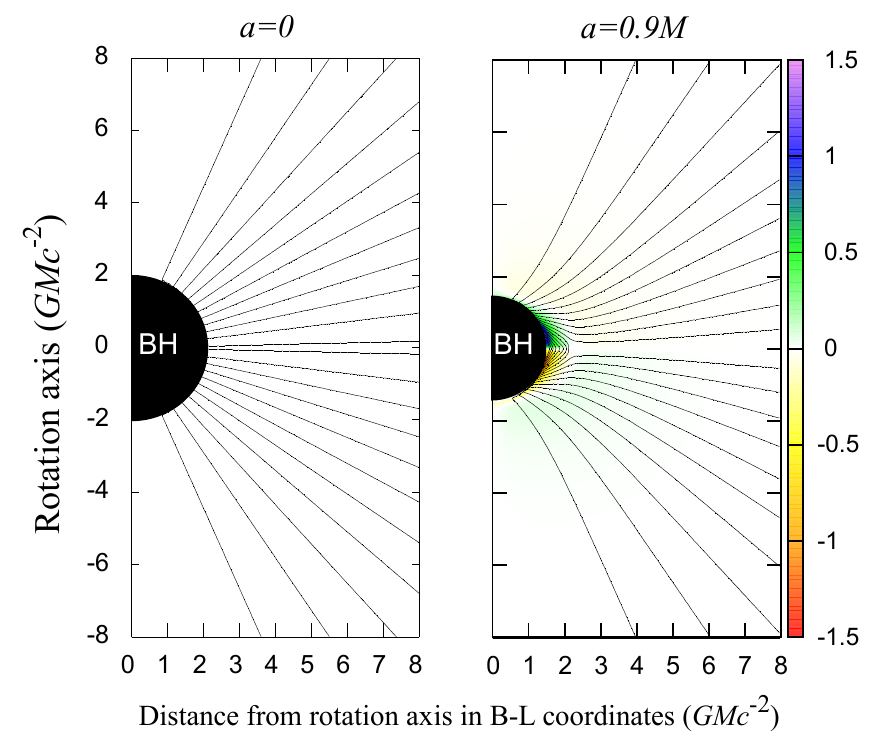}
\caption{
Stationary equi-$A_\varphi$ contours (solid curves) and 
the distribution of 
$\mbox{\boldmath$E$} \cdot \mbox{\boldmath$B$} / B_{\rm eq}(2M){}^2$
(color) on the poloidal plane ($r$,$\theta$).
The equitorial current density is assumed to depend on $r$ as
$J^\varphi{}_{\rm eq} \propto r^{-4}$.
The ordinate represents the distance
along the rotational axis of the BH, 
while the abscissa does the distance $r\sin\theta$ 
from the rotation axis,
Both axes show the lengths
in the Boyer-Lindquist coordinates,
normalized by the gravitational
radius, $r_{\rm g}=GMc^{-2}=M$.
The equatorial plane corresponds to
the ordinate of $0$.
The black filled circle shows the BH.
Left panel shows the equi-$A_\varphi$ contours when $a=0$,
in which case the solid curves denote the magnetic field lines measured by a distant static observer.
For $a=0$, there arises no electric fields along the magnetic field lines;
thus, the background color is entirely white.
Right panel shows the case of $a=0.9M$.
The amplitude of 
$\mbox{\boldmath$E$} \cdot \mbox{\boldmath$B$}$ 
increases with increasing $a$,
because the frame-dragging effect increases
with increasing BH's spin, $a$.
            }
    \label{fig:Epara_0}
\end{figure*}

In the present paper, we consider a ten solar-mass BH,
$M_1 \equiv M/(10M_\odot)=1$,
and solve equations~(\ref{eq:BS_3}) and 
(\ref{eq:Poisson_4}) iteratively in the region
$r_{\rm H} < r \le 20M$ and
$0 \le \theta \le \pi$.
The solved electromagnetic fields are presented
in figure~\ref{fig:Epara_0}.
In each panel, thin black curves show the equi-$A_\varphi$ contours,
which indicate the poloidal magnetic field lines
for a distant static observer if $a=0$.
We superpose
$ \mbox{\boldmath$E$} \cdot \mbox{\boldmath$B$}
 / [B_{\rm eq}(r=2M)]^2
$,
whose values are indicated in the color code.
The left and right panels correspond to the cases of
$a=0$ and $a=0.9M$, respectively.

In the left panel, 
$ \mbox{\boldmath$E$} \cdot \mbox{\boldmath$B$} = 0 $
holds everywhere; 
thus, the background color is entirely white.
It follows that
the poloidal magnetic field becomes radial 
for a slowly rotating BH when 
$J^\varphi{}_{\rm eq} \propto r^{-4}$,
which is consistent with the analytical conclusion
\citep{bla77,mck07}.
Because of $a=0$, 
there exists no magnetic-field-aligned 
electric field, $E_\parallel$;
thus, $A_\varphi$ is solved 
only from the Biot-Savart law.

\begin{figure*}
\vspace*{-0.0truecm}
\hspace{3.8cm}
\includegraphics[width=0.6\textwidth]{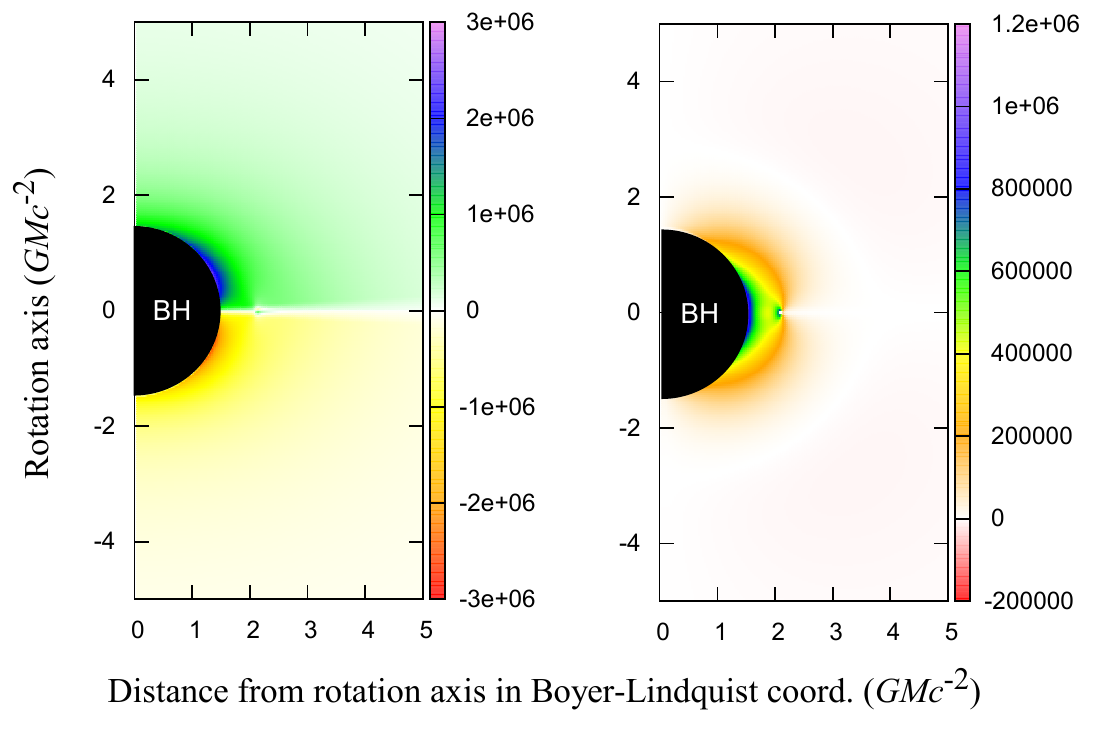}
\vspace*{-0.0truecm}
\caption{
Poloidal magnetic field measured in ZAMO at $t=0$ for $a=0.9M$.
The abscissa and ordinate are common with
figure~\ref{fig:Epara_0},
but the BH vicinity is closed up.
Left and right panels show $B^{\hat r}({\rm ZAMO})$ and
$B^{\hat{\theta}}({\rm ZAMO})$, respectively, in Gauss.
            }
    \label{fig:Bp}
    
\end{figure*}

However, as the BH spins up (i.e., if $a \ne 0$),
the right panel shows that
there arises a non-vanishing $E_\parallel$
due to the relative rotational motion of the magnetic field lines
with respect to the space time.
At the same time, equi-$A_\varphi$ lines deviates from 
a radial shape, as the solid curves indicate.
For a rapidly rotating case, $a=0.9M$,
the magnetic field lines are laterally pushed 
toward the rotation axis in the ergosphere
\citep{Tanabe:2008:PhRvD,Tchekhovskoy:2010:ApJ}.
Note that the magnetosphere is assumed to be vacuum in the present analysis,
while it is force-free (i.e., plasma-filled) in
\citet{bla77,Tanabe:2008:PhRvD,Tchekhovskoy:2010:ApJ}.
To examine how the magnetic field lines are actually deformed
for a physical observer, we adopt ZAMO and plot in figure~\ref{fig:Bp}
the radial ($\tilde{B}^{\hat{r}}$, left panel) and 
the meridional ($\tilde{B}^{\hat\theta}$, right panel) 
components of the magnetic field,
where their explicit expressions are given by
equations~(\ref{eq:Br_ZAMO}) and (\ref{eq:Bth_ZAMO}).
It shows that $\vert \tilde{B}^{\hat{\theta}} \vert$ 
is kept below
$\vert \tilde{B}^{\hat{r}} \vert$ 
except for the equatorial region
where $\tilde{B}^{\hat{r}}$ vanishes by symmetry.
To compare with slowly rotating cases, 
we adopt the same power-law 
in the disk current, 
$J^\varphi{}_{\rm eq} \propto r^{-4}$,
for all the cases of $a$.
To avoid a substantial 
$\vert \tilde{B}^{\hat{\theta}} \vert$ 
in the lower-latitude ergosphere,
we could adopt another functional form 
of $J^\varphi{}_{\rm eq}$ for $a \ne 0$;
however, such a fine-tuning 
is out of the scope of the present paper.

We assume a positive $J^\varphi{}_{\rm eq}$; 
thus, $F_{\theta\varphi}$
(i.e., the radial component of the magnetic field) 
is positive (or negative) in the northern (or southern) hemisphere.
Accordingly, a positive (or a negative) sign of 
$\mbox{\boldmath$E$} \cdot \mbox{\boldmath$B$}$ indicates 
that the electric field points outwards 
in the northern (or southern) hemisphere.
Thus, as plasmas are created (via photon-photon collisions) 
near the horizon, $r<2M$, 
electrons (or positrons) are accelerated inwards (or outwards) 
in both hemispheres
in this stationary solution.
Such accelerated electrons and positrons produce electric currents
in the magnetosphere, whose poloidal components
modifies the poloidal electric field through the Ampere's law.
In the next section, we will focus on 
the temporal evolution of the electromagnetic fields and
the particle distribution functions starting from the initial conditions described in this section.


\section{The particle-in-cell (PIC) scheme}
\label{sec:PIC}
Let us look briefly at the collisionless nature of the plasmas
in \S~\ref{sec:collisionless},
before turning to a closer examination of the temporal evolution
of the BH magnetosphere in the rest of this section.

\subsection{Collisionless plasmas}
\label{sec:collisionless}
Denoting the density of a pair plasma with 
$n_\pm = \kappa n_{\rm GJ}$, 
we can express the collision frequency as
\begin{equation}
  \nu_{\rm c} \sim \kappa n_{\rm GJ} \sigma c,
  \label{eq:nu_c1}    
\end{equation}
where 
$\sigma$ refers to the collisional cross section, and
\begin{equation}
  n_{\rm GJ} \equiv \omega_{\rm H} B / (4\pi ce)
  \label{eq:nGJ}
\end{equation}
denotes the Goldreich-Julian (GJ) number density,
which is rotaionally induced;
$e$ denotes the charge on the electron.
If the plasma density is comparable to the GJ value,
$\kappa$ becomes of the order of unity.

Let us evaluate the cross section by
$\sigma \sim \pi l_{\rm c}{}^2$,
where $l_{\rm c}$ denotes the typical impact parameter.
Equating the potential and kinetic energies, we find
$e^2/l_{\rm c} \sim (\gamma-1) m_{\rm e} c^2$.
Then combining it with equation~(\ref{eq:nu_c1}), we obtain
\begin{equation}
  \nu_{\rm c} \sim \frac{\kappa \pi n_{\rm GJ} e^4}
                        {\gamma^2 m_{\rm e}{}^2 c^3}
  \label{eq:nu_c2},
\end{equation}
where $\gamma \gg 1$ is assumed.
On the other hand, the gyro frequency is given by
\begin{equation}
  \nu_{\rm B}= \frac{e B}{2\pi \gamma m_{\rm e} c}.
  \label{eq:nu_B}
\end{equation}
We thus obtain
\begin{equation}
  \frac{\nu_{\rm c}}{\nu_{\rm B}}
  \sim \frac{\pi}{4} \frac{\kappa}{\gamma}
       \frac{a}{M} \frac{r_0}{r_{\rm H}},
  \label{eq:collisionless}
\end{equation}
where $r_0 \equiv e^2/(m_{\rm e} c^2)$ denotes the classical 
electron radius.
Since $r_0 / r_{\rm H} \sim 10^{-19} M_1{}^{-1}$ holds,
we can conclude that the collision frequency 
is much less than the gyro frequency,
and that the assumption of collisionless plasmas in 
the PIC scheme is justified.
This conclusion solely comes from the fact
that the GJ density corresponds to a high vacuum
in BH or pulsar magnetospheres.

It should be noted that the Ohm's law,
which is necessary to close the system of equations in MHD,
requires that many collisions should take place within a single gyration.
However, equation~(\ref{eq:collisionless}) shows 
this assumption cannot be justified in a black hole magnetosphere
unless the plasma density greatly exceeds the GJ value.
In the present paper, we thus construct the electric current 
from the actual motion of charged particles, adopting the PIC scheme.
By this method, 
we can, for instance, 
incorporate the currents carried by the drift motion
of charged particles,
as well as the anisotropic distribution of particles' momenta 
(e.g., along and perpendicular to the magnetic field lines).

\subsection[]{The Maxwell equations}
\label{sec:Maxwell}
In the present paper, we assume $F_{\varphi t}=0$
throughout the simulations.
Accordingly, together with $\partial_\varphi=0$ for all the quantities,
we find
$\partial_t F_{\theta\varphi}=0$ and
$\partial_t F_{\varphi r}=0$
from the Faraday's law.
Thus, in the present paper, we treat both $B^r$ and $B^\theta$ are
constant in time (and hence unchanged from the initial condition), 
and solve the temporal evolution of 
only $F_{r t}$, $F_{\theta t}$, and $F^{r \theta}$.
This assumption of $F_{\varphi t}=0$ is 
justifined as long as
the toroidal currents curried by the simulated 
particles are small compared to the stationary, equatorial toroidal current
that is carried by the accreting plasmas.

Under this assumption,
the Faraday and Ampere's laws give
\begin{equation}
  \frac{\partial F^{r \theta}}
       {\partial t}
  = -\frac{\Delta}{\Sigma^2}
     \left( \frac{\partial F_{\theta t}}{\partial r}
           -\frac{\partial F_{r t}}{\partial \theta}
     \right),
  \label{eq:Maxwell_1}
\end{equation}
\begin{equation}
  \frac{\partial F_{r t}}
       {\partial t}
  = \frac{\Sigma^2}{A}
    \left[ \frac{1}{\sqrt{-g}}
           \frac{\partial}{\partial \theta}
           \left( \sqrt{-g} F^{r \theta} \right)
          -4\pi J^r
    \right],
  \label{eq:Maxwell_2}
\end{equation}
\begin{equation}
  \frac{\partial F_{\theta t}}
       {\partial t}
  = \frac{\Sigma^2 \Delta}{A}
    \left[-\frac{1}{\sqrt{-g}}
           \frac{\partial}{\partial r}
           \left( \sqrt{-g} F^{r \theta} \right)
          -4\pi J^\theta
    \right].
  \label{eq:Maxwell_3}
\end{equation}
Replacing the $r$ derivatives with $r_\ast$ derivatives,
and introducing the following dependent variables,
\begin{eqnarray}
  B
  && 
  \equiv \sqrt{-g} F^{r \theta},
  \nonumber\\
  D
  && 
  \equiv F_{r t},
  \nonumber\\
  E
  && 
  \equiv F_{\theta t} \sin\theta,
  \label{eq:def_EDB}
\end{eqnarray}
all of which are well-behaved at the horizon,
we can rewrite the three 
Maxwell equations~(\ref{eq:Maxwell_1})--(\ref{eq:Maxwell_3}) into
\begin{equation}
  \frac{\partial B}{\partial t}
  = -c_1 \frac{\partial E}{\partial x}
    +c_2 \frac{\partial D}{\partial y}
  \label{eq:Maxwell_1b}
\end{equation}
\begin{equation}
  \frac{\partial D}{\partial t}
  =  c_3 \frac{\partial B}{\partial y}
    -4\pi \frac{\Sigma^2}{A} J^r
  \label{eq:Maxwell_2b}
\end{equation}
\begin{equation}
  \frac{\partial E}{\partial t}
  = -c_4 \frac{\partial B}{\partial x}
    -4\pi \frac{\Sigma^2}{A} \Delta\sin\theta \cdot J^\theta
  \label{eq:Maxwell_3b}
\end{equation}
where 
\begin{equation}
  x \equiv r_\ast, \qquad 
  y \equiv 1-\cos\theta,
\end{equation}
\begin{equation}
  c_1 \equiv \frac{r^2+a^2}{\Sigma}, \qquad
  c_2 \equiv \frac{\Delta \sin^2\theta}{\Sigma},
  \nonumber
\end{equation}
\begin{equation}
  c_3 \equiv \frac{\Sigma}{A}, \qquad
  c_4 \equiv \frac{\Sigma (r^2+a^2)}{A}.
  \label{eq:app_c4}
\end{equation}
All the coefficients $c_1$, $c_2$, $c_3$, and $c_4$ are 
positive definite. 
Both $c_1$ and $c_4$ are close to unity in the entire region.
However, $c_2 \ll 1$ holds near the horizon or near the pole
but $c_2 \rightarrow 1$ at $r \gg M$ in the lower latitudes.
We have $c_3 \ll 1$ at $r \gg M$ and $c_3 \approx 0.25$ near the horizon.
Note that the singular behaviour (i.e., the polynomial pole) 
at $\theta=0$ 
in equations~(\ref{eq:Maxwell_2}) and (\ref{eq:Maxwell_3})
is eliminated by introducing a new meridional coordinate, $y$.


To solve these three Maxwell equations
(eqs.~[\ref{eq:Maxwell_1b}]--[\ref{eq:Maxwell_3b}]),
we must impose appropriate boundary conditions
that are consistent with the initial, stationary state
(\S\ref{sec:BDCs}).
Along the northern and southern polar axes 
(i.e., at $\theta=0$ and $\theta=\pi$),
we impose
\begin{equation}
  B = 0, \quad
  \frac{\partial D}{\partial y} = 0, \quad
  E = 0.
  \label{eq:BDC_1}
\end{equation}
At the inner boundary, $x=r_\ast=-\infty$, we impose
\begin{eqnarray}
  \frac{\partial B}{\partial x} = 0, & & \quad
  \tilde{E_{\hat{r}}} \propto D+\omega F_{r \varphi}=0,  \nonumber \\
          & &\frac{\partial E}{\partial x} 
  = \left( \frac{\partial E}{\partial x}
    \right)_{t=0},
  \label{eq:BDC_3}
\end{eqnarray}
where the quantity within $()_{t=0}$ indicates the 
initial value at $t=0$.
At the outer boundary, $x=r_\ast= r_{\rm out}$, we impose
\begin{equation}
  \frac{\partial B}{\partial x} = 0, \quad
  D = 0, \quad
  \frac{\partial E}{\partial x} 
  = \left( \frac{\partial E}{\partial x}
    \right)_{t=0}.
  \label{eq:BDC_4}
\end{equation}
Note that we set $B=0$ in the entire region at $t=0$.

\subsection[]{Particle equation of motion}
\label{sec:EOM}
In a highly vacuum BH magnetosphere,
charged leptons are deccelerated by 
the radiation-reaction forces.
To include these forces from the first principles,
we must adopt tiny time steps 
and consider the force on one part of the charge
by the fields of another part,
taking account of retardation within the particle itself.
However, in actual simulations, 
it is unrealistic to adopt such tiny timesteps.
Thus, as a compromize, we include the radiation reaction force
as a friction term in the equation of motion (EOM).

With such a friction term, the EOM can be expressed as
\citep{Thorne:1982:MNRAS,Hughes:1994:PhRvD,Bacchini:2018:ApJS}
\begin{eqnarray}
  \frac{du_i}{dt}
  = 
  &-&\alpha u^t \partial_i \alpha 
    +u_k \partial_i \beta^k
    -\frac12 \frac{u_j u_k}{u^t} \partial_i g^{jk}
  \nonumber\\
  &+&
     \frac{q}{m} \left( F_{it}+F_{ij}\frac{u^j}{u^t} \right)
    +\frac{(F_{\rm rad})_i}{u^t},
  \label{eq:EOM}
\end{eqnarray}
where $\alpha$ is defined by equation~(\ref{eq:lapse}), 
$\beta^r=\beta^\theta=0$, $\beta^\varphi= g_{t\varphi}/g_{\varphi\varphi}$,
$q/m$ refers to the charge-to-mass ratio, and
\begin{equation}
  u^t= \frac{\sqrt{1+g^{jk} u_j u_k}}{\alpha}.
  \label{eq:ut}
\end{equation}
The Latin indices run over 1, 2, 3.
We can evaluate the radiation-reaction force by
the covariant form,
\citep[\S~17.8 of][]{jackson62}
\begin{equation}
  F_{\rm rad}^j
  = \frac{2}{3} \frac{r_0}{c} 
    \left( \frac{d^2 u^j}{d\lambda^2}
          +u^j \frac{du^\nu}{d\lambda} \frac{du_\nu}{d\lambda}
    \right),
  \label{eq:def_Prad}
\end{equation} 
where $r_0$ denotes the classical electron radius, 
$\lambda$ does the particle's proper time,
and the Greek indices run over 0, 1, 2, 3.
This radiation reaction force includes the effects of any kind of
acceleration acting on the particles.
For instance, photon emissions as a result of the acceleration
in an electro-magnetic field (e.g., via the synchro-curvature process)
and that in a gravitational field, are included
(Appendix~\ref{sec:app_covaniance}).
However, this radiation-reaction force does not include
the effect of inverse-Compton scatterings,
which should be considered separately in future works.

%

Definition of the four velocity $u^\mu$ gives
\begin{equation}
  \frac{dr}{dt}
  = \frac{u^r}{u^t},
  \label{eq:drdt}
\end{equation}
\begin{equation}
  \frac{d\theta}{dt}
  = \frac{u^\theta}{u^t},
  \label{eq:dthdt}
\end{equation}
and
\begin{equation}
  \frac{d\varphi}{dt}
  = \frac{u^\varphi}{u^t}.
  \label{eq:dphdt}
\end{equation}

For presentation purpose, we can convert $u_i= dx_i/d\lambda$ 
in terms of the ZAMO-measured spatial velocity, $u_{\hat j}$,
as described in Appendix~\ref{sec:app_ZAMO}.


We integrate both equations~(\ref{eq:EOM}) and
(\ref{eq:drdt})--(\ref{eq:dphdt}) in the phase space 
with the global time variable $t$,
which corresponds to the proper time of a distant static observer
(i.e., us).

Let us briefly describe the boundary conditions on
the motion of electrons and positrons.
Due to the symmetry, 
we assume that the particles moving across the polar axis
(at $\theta=0$ or $\pi$) will be reflected 
equator-ward with opposite meridional velocity.
Both the inner and outer boundaries are treated as particle sinks.
Thus, when particles move accross these two radial boundaries,
they are excluded from the simulation.

\subsection[]{Plasma supply}
\label{sec:supply}
In BH magnetospheres, pairs can be supplied via 
two-photon and/or one-photon (i.e., magnetic) 
pair production processes.
In the present paper, we focus on 
the former process, and consider the collisions
of MeV photons emitted from the equatorial ADAF
via Bremsstrahlung.
In subsequent papers, we will also consider
the collisions between the gap-emitted (inverse-Compton)
photons and the ADAF-emitted (synchrotron) photons.

The pair supply rate (pairs per second per volume)
is given by
\begin{equation}
    \dot{N}_\pm = c \sigma_{\gamma\gamma} n_\gamma{}^2,
    \label{eq:dotN}
\end{equation}
where $\sigma_{\gamma\gamma}$ denotes the total cross section 
of photon-photon pair production, and 
$n_\gamma$ does the MeV photon density.
Adopting the Newtonian self-similar ADAF model
\citep{Mahadevan:1997:ApJ},
and assuming that the most energetic MeV photons
are emitted within $r=4M$,
we obtain
(Appendix~\ref{sec:app_ADAF})
\begin{equation}
  \dot{N}_\pm
  \approx 1.0 \times 10^{24}
    \dot{m}^4 M_1{}^{-2}
    \max \left[ \left(\frac{r}{4M}\right)^{-4}, 1 \right]
  \label{eq:dotN2}
\end{equation}
in cgs unit (i.e., in $\mbox{pairs s}^{-1}\mbox{ cm}^{-3}$).

We randomly introduce a macro particle in each cell 
at every time step with probability 
$1/k_{\rm create}=0.1$;
that is, particles are injected in each cell 
at every $k_{\rm create}=10$ time steps on average.
In this case, each created macro positron or electron
has the electric charge
\begin{equation}
    q_i= \pm e \dot{N}_\pm k_{\rm create} \Delta_t \Delta_V,
\end{equation}
where $\Delta_t$ denotes the interval of each time step, and
$\Delta_V$ the invariant volume of each cell.
Note that 
$\Delta_t \Delta_V= \sqrt{-g} dt dr d\theta d\varphi
 = 2\pi \sqrt{-g} \Delta_t \Delta_r \Delta_\theta$
holds, 
where $\Delta_r$ and $\Delta_\theta$ denote
the intervals in Boyer-Linquist 
radial and meridional coordinates, 
both of which are non-uniformly gridded.

In the initial state, 
there are no macro particles in any cell.
As the PIC simulation proceeds, 
the number of macro particles
increases with $t$ to saturate at a few hundred 
in each PIC cell on average.
Here, the maximum value of the Courant number is set
to be $0.5$ for uniform grid intervals 
in $x=r_\ast$ and $y=1-\cos\theta$ coordinates.
In total, there are about $3 \times 10^8$ 
macro particles in the entire simulation region.

\subsection[]{Nonstationary magnetosphere}
\label{sec:nonstationary}
It is checked {\it a posteriori} 
(\S~\ref{sec:disc_resolution})
that the invariant grid intervals resolve 
the skin depth
\begin{equation}
 l_{\rm p}= \frac{c}{\omega_{\rm p}},
 \label{eq:skin_depth}
\end{equation}
at every point at any elapsed time,
where the plasma frequency $\omega_{\rm p}$ is computed by the plasma density
$n_\pm$ and its mean Lorentz factor $\langle\gamma\rangle$ as
\begin{equation}
  \omega_{\rm p}= \sqrt{\frac{4\pi e^2 n_\pm}{m_{\rm e} \langle \gamma \rangle}}.
  \label{eq:plasma_freq}
\end{equation}

For stellar-mass BHs, we obtain
\begin{equation}
  \frac{l_{\rm p}}{r_{\rm g}}
  = 
    6.84 \times 10^{-2} k
   \gamma_6{}^{1/2}
   \left( \frac{a}{M} \cdot \frac{B}{B_{\rm eq}} \right)^{-1/2}
   M_1{}^{-1/4},
   \label{eq:sdepth_2}
\end{equation}
where $\gamma_6 \equiv \langle \gamma \rangle / 10^6$, and
\begin{equation}
  k 
  \equiv 
  \kappa^{-1/2}
  \dot{m}^{-1/4}
  \left( \frac{r}{r_{\rm H}} \right)^{5/8}
  \left( \frac{m_{\rm e}}{m_{\rm p}/1836} \right)^{1/2};
\end{equation}
$m_{\rm p}$ denotes the rest mass of the proton.

To resolve the skin depth, $l_{\rm p}$,
most PIC simulations are performed in one or two dimensions
with small ion-to-electron mass ratios
(e.g., \citet{Bohdan:2019:ApJ}),
covering limited time and length scales.
In the present paper,
adopting a heavy electron mass of
$m_{\rm e}= m_{\rm p}/20$,
we perform a two-dimensional PIC simulation
within about ten gravitational radii
over the time duration of several hundred dynamical time scales.
We rescale every expression according to this modified electron mass.
Under this assumption of heavy charged leptons,
we need about $10^3$
grid points to fully resolve the skin depth.

On these grounds,
we adopt a radial grid of 960 uniform cells between
$-15.81807 < x < 12.82443$,
which corresponds to
$1.46514M < r < 13.68548M$,
and 1920 uniform cells between $0 < y=1-\cos\theta < 2$,
which corresponds to $0^\circ < \theta < 180^\circ$.
We present the explicit expression of the discretized
Maxwell equations and the particle equation of motion
in Appendix~\ref{sec:discretization}.
To construct the electric currents from particle motion
(Appendix~\ref{sec:current}),
we adopt the area weighting \citep{villa92}.

\subsubsection{Electromagnetic fields}
\label{sec:acc}
\begin{figure*}[t]
\vspace*{-0.0truecm}
\hspace{2.8cm}
\includegraphics[width=0.7\textwidth]{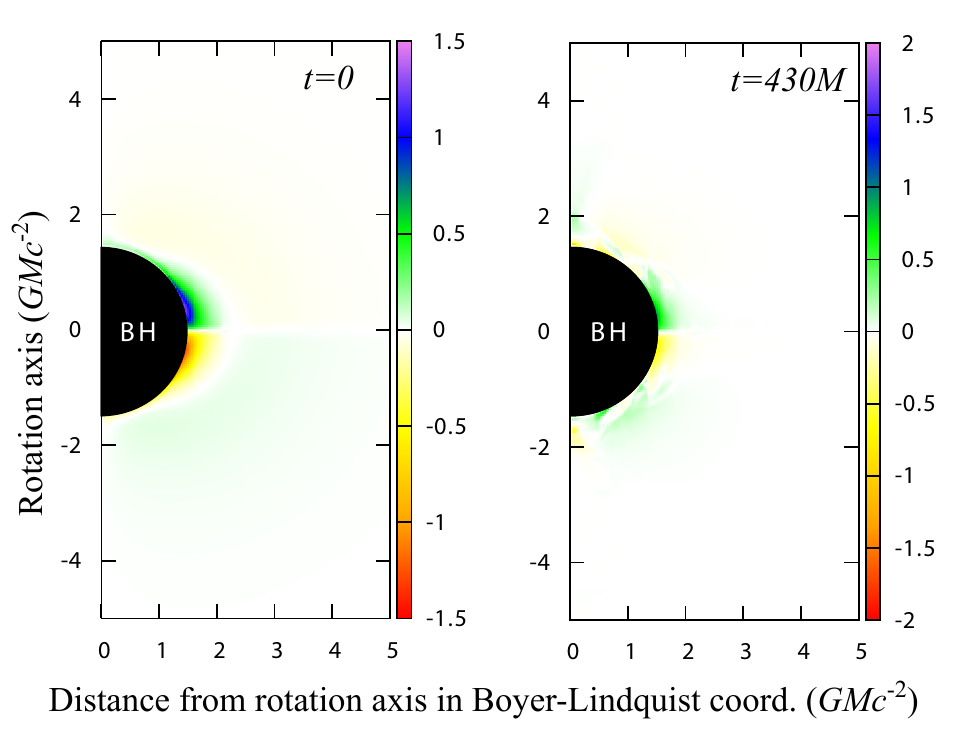}
\vspace*{-.0truecm}
\caption{
Distribution of 
$\mbox{\boldmath$E$} \cdot \mbox{\boldmath$B$} 
 / B_{\rm eq}(2M){}^2$
on the poloidal plane ($r$,$\theta$)
for $a=0.9M$, $B=B_{\rm eq}$, and $\dot{m}=2.25 \times 10^{-4}$.
Left and right panels show the distribution at 
$t=0.00$ and $t=430.00 GM c^{-3}$, respectively.
The abscissa and ordinate are common with
figure~\ref{fig:Bp}.
Magnetic surfaces (i.e., constant $A_\varphi$ lines) 
are not depicted, but are common with
the right panel of figure~\ref{fig:Epara_0}
            }
    \label{fig:Epara_1}
\end{figure*}

We start with the magnetic-field-aligned electric field.
In figure~\ref{fig:Epara_1}, 
we present the distribution of 
$ \mbox{\boldmath$E$} \cdot \mbox{\boldmath$B$} / [B_{\rm eq}(2M)]^2$.
For $M=10 M_\odot$ and 
$\dot{m}=2.25 \times 10^{-4}$, 
we obtain
$B_{\rm eq}(2M)= 1.452 \times 10^{6}$~G.
The left and right  panels show the distribution of 
$ \mbox{\boldmath$E$} \cdot \mbox{\boldmath$B$} / [B_{\rm eq}(2M)]^2$ 
at $t=0$ and $t=430 M$, respectively.
In the northern hemisphere, 
since $F_{\theta\varphi}>0$ holds, 
a positive (or a negative) sign of 
$\mbox{\boldmath$E$} \cdot \mbox{\boldmath$B$}$
indicates that an outward (or an inward) electric field arises
along the local magnetic field lines.
In the same manner, in the southern hemisphere,
it follows from $F_{\theta\varphi}<0$ that
a positive (or negative) sign of 
$\mbox{\boldmath$E$} \cdot \mbox{\boldmath$B$}$
means an inward (or an outward) electric field.

\begin{figure*}[t]
\hspace{0.0cm}
\includegraphics[width=1.0\textwidth, angle=0]{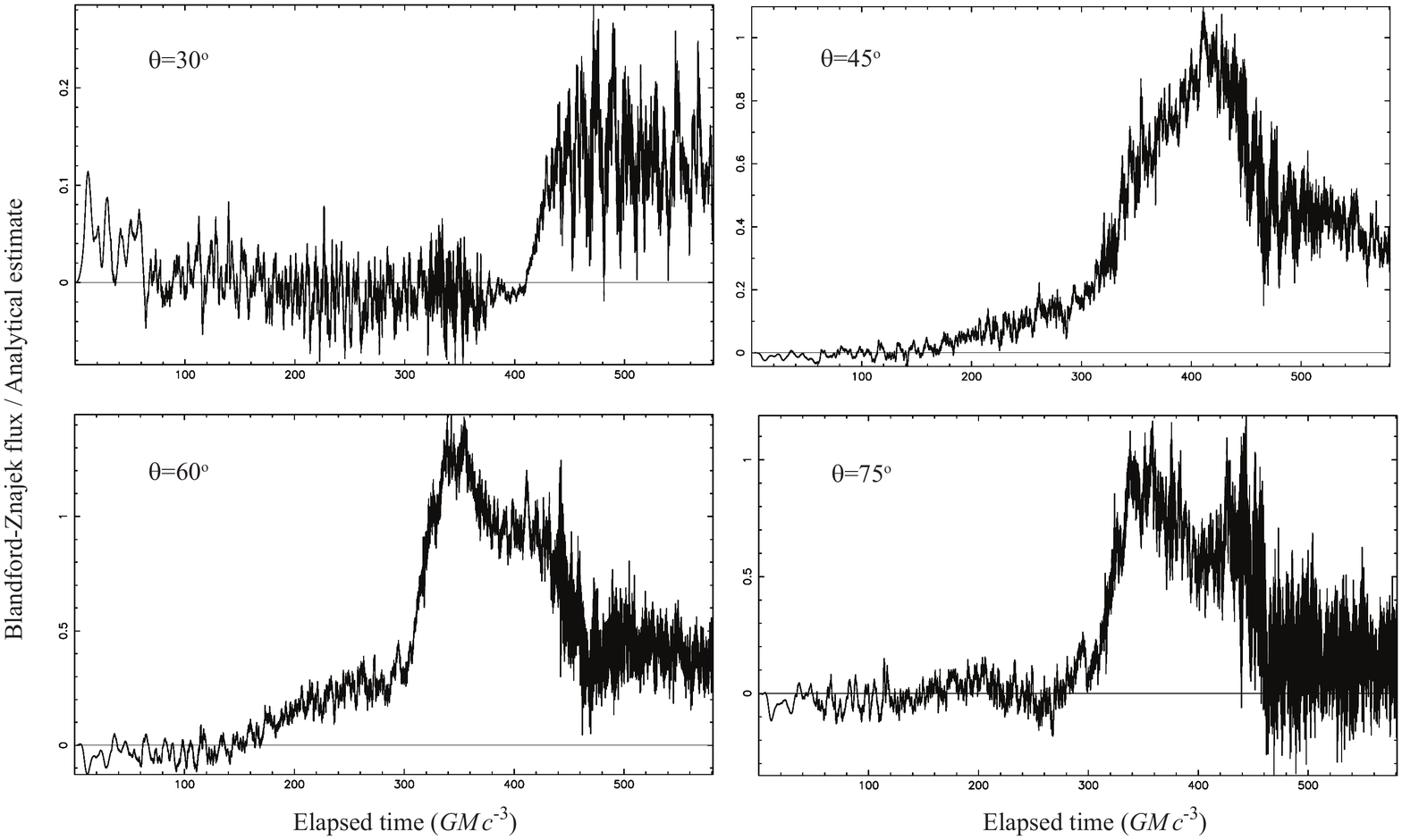}
\vspace*{-2.0truecm}
\caption{
Radial component of the Blandford-Znajek (BZ) flux 
as a function of the elapsed time at radius $r=r_{\rm H}+0.25M$.
The ordinate is normalized by its analytical estimate
(see text for details),
while the abscissa by the dynamical time scale.
Each panel shows the BZ flux at discrete colatitudes as labeled.
}
    \label{fig:LBZ_Time}
\end{figure*}

At $t=0$,
there are no poloidal currents in the magnetosphere
because of the lack of plasmas.
Thus, the magnetic field has no toroidal component. 
Accordingly, the rotational energy of the BH is not 
being extracted.
On the other hand, because of the relative motion
of the magnetic field lines with respect to 
the space time, 
there appears a strong electromagnetic 
field along the mangetic field line.
The left panel of figure~\ref{fig:Epara_1}
shows that 
such a magnetic-field-aligned electric field 
is exerted in both hemispheres.

As time elapses, 
$\vert \mbox{\boldmath$E$} \cdot \mbox{\boldmath$B$} \vert$
evolves, exerting electric currents in the magnetosphere.
These currents modify the poloidal electric field through the Ampere's law.
For example, at $t=430M$ in the northern hemisphere,
the right panel shows that a negative
$\mbox{\boldmath$E$} \cdot \mbox{\boldmath$B$}$ 
appears in the higher-middle latitudes,
exerting inwards currents there.
In the same manner, in the southern hemisphere, 
$\mbox{\boldmath$E$}$ points inwards
in the higher-middle latitudes outside the ergosphere.
Note that $F_{\theta\varphi}$ changes sign
across the equator.

Because of this nearly symmetric distribution of the electric field
between the two hemispheres,
electrons (or positrons) are accelerated outwards (or inwards)
in the higher-middle latitudes in both hemispheres.
In the lower latitudes, however,
positrons (or electrons) are accelerated outwards (or inwards)
by the strong magnetic-field-aligned electric field
in both hemispheres.
As a result, currents flow inwards in the middle latitudes
and flow outwards in the lower latitudes.
For the details of the electric currents, 
see \S~\ref{sec:ptcl}.
Such a pattern of the magnetospheric currents lead to 
a positive toroidal magnetic field 
($\propto F^{r \theta}$)
in the northern hemisphere, 
while a negative $F^{r \theta}$ in the southern hemisphere.
Thus, $F^{r \theta}$ vanishes on the equator.

Now let us consider the Poynting flux,
or equivalently the BZ flux,
\begin{equation}
    T_{\rm em}{}^r{}_t = \frac{c}{4\pi} F^{r \mu} F_{\mu t}
            \propto F^{r \theta} F_{\theta t}
  \label{eq:Poynting}
\end{equation}
in the BH vicinity.
If it becomes positive,
it means that the BH's rotational energy 
is being extracted electromagnetically.
In figure~\ref{fig:LBZ_Time},
we plot the temporal evolution of 
$T_{\rm em}{}^r{}_t (t,r,\theta)$
at $r= r_{\rm H}+0.25M$
at four discrete colatitudes as labeled.
The ordinate is normalized by the typical BZ flux, 
which is analytically given by
$F_{\rm analytical} \equiv L_{\rm BZ}/S_{\rm area}$,
where 
$S_{\rm area} \equiv \int\int \sqrt{A}\sin\theta d\theta d\phi$.
The BZ power (i.e., the spin-down luminosity)
can be estimated to be
\begin{equation}
  L_{\rm BZ}
  = \frac{1}{128}
    \left( \frac{a}{M} \right)^2
     B_\perp{}^2 r_{\rm H}{}^2 c
  \label{eq:LBZ}
\end{equation}
in the slow-rotating limit, $\vert a \vert \ll M$,
where $B_\perp$ denotes the averaged strength of 
the radial magnetic field.
It follows from the figure that the solution exhibits
rapid plasma oscillations,
as reported by \citet{Levinson:2018:AA}.
It also follows that the simulated BZ flux is consistent with 
its analytical estimate.
See the supplementary material of
\citet{Crinquand:2020:PhRvL}
for a consistent discussion between the simulated BZ flux
and their analytical estimates.

It should be noted that the BZ flux increases during the elapsed time
$ 320M < t < 450M$ along the middle latitudes,
$45^\circ \le \theta \le 75^\circ$.
During this flux-enhancement phase,
the BZ flux (i.e., the Poynting flux) fluctuates relatively mildly 
compared to its amplitude
within $45^\circ \le \theta \le 60^\circ$.
In the present magnetically dominated magnetosphere,
in which the magnetic energy density dominates
the particles' rest-mass energy densities,
particles' energy flux is typically less than 
$10^{-7}$ compared to the Poynting flux; 
thus, we neglect their contribution 
when we consider the energy flux.

\begin{figure*}[t!]
\hspace{1.2cm}
\includegraphics[width=0.8\textwidth, angle=0]{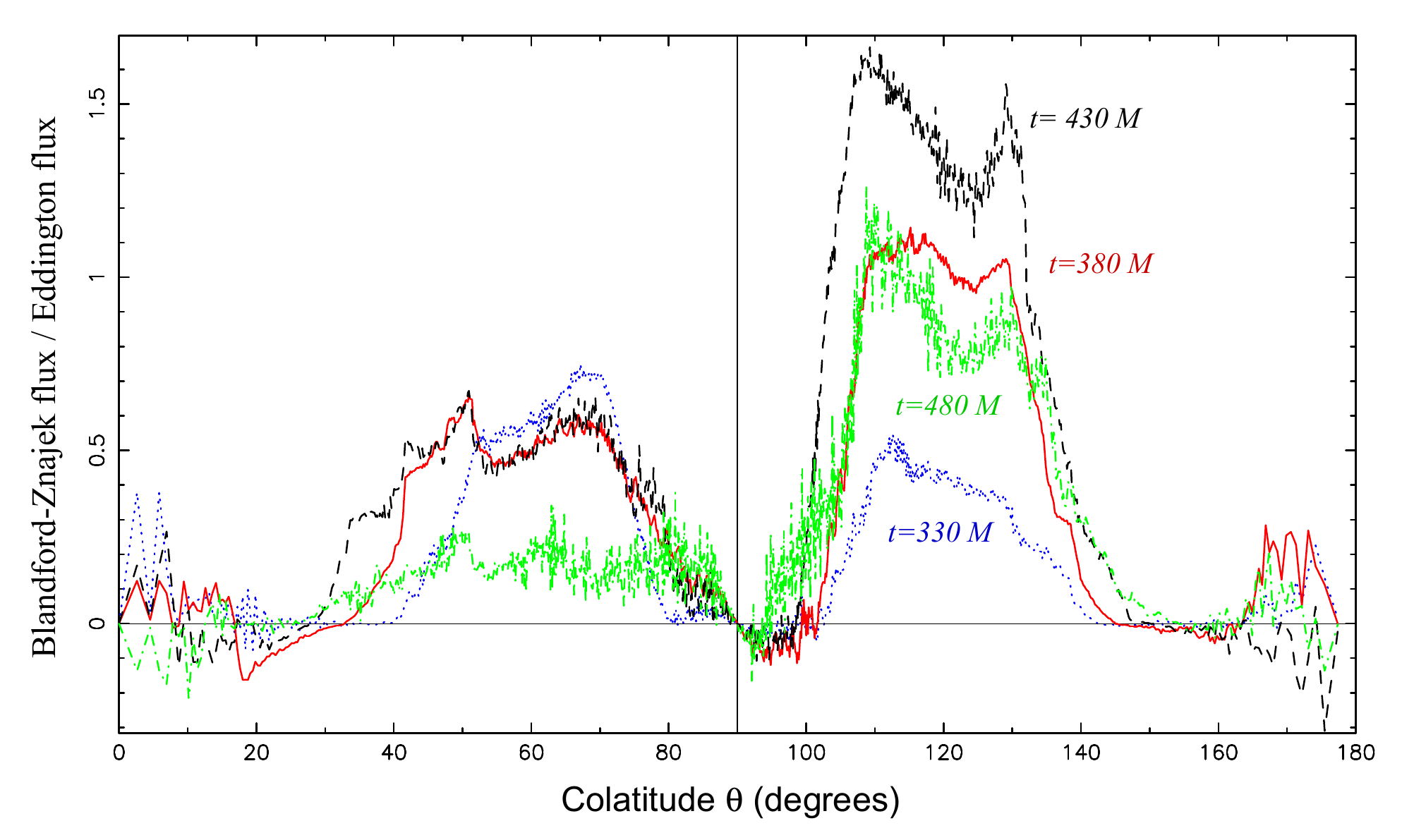}
\vspace*{-.truecm}
\caption{
Radial component of the Blandford-Znajek flux 
as a function of the colatitude angle, $\theta$.
The blue dotted, red solid, black dashed, and green dash-dotted 
curves show the BZ flux at time
$t=330.00 GMc^{-3}$, $380.00 GMc^{-3}$, $430.00 GMc^{-3}$, 
and $480.00 GMc^{-3}$, respectively.
            }
    \label{fig:LBZ_Angle}
\end{figure*}

In figure~\ref{fig:LBZ_Angle},
we present the angular dependence of the BZ flux
at four elapsed time as labeled.
During the flux-enhancement phase,
the BH's rotational energy is efficiently extracted
from the middle latitudes,
$40^\circ < \theta < 75^\circ$ in the northern hemisphere,
and 
$105^\circ > \theta > 135^\circ$ in the southern hemisphere.

\begin{figure*}[t!]
\hspace{0.0cm}
\includegraphics[width=1.0\textwidth, angle=0]{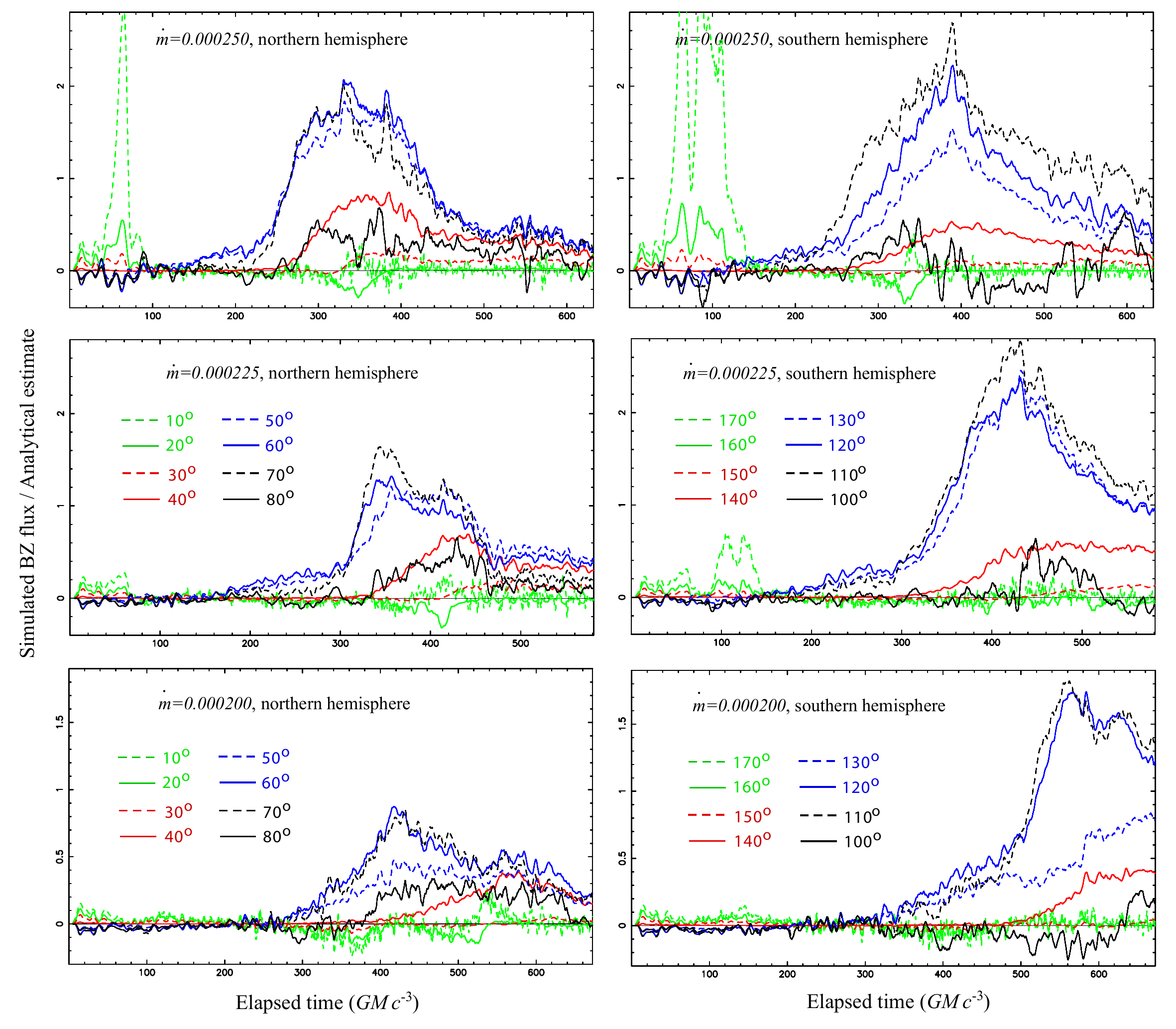}
\vspace*{-.5truecm}
\caption{
Moving-averaged BZ fluxes with a period of $5 GMc^{-3}= 5M$,
for three dimensionless accretion rates,
$\dot{m}=2.50 \times 10^{-4}$ (top),
$\dot{m}=2.25 \times 10^{-4}$ (middle), and
$\dot{m}=2.00 \times 10^{-4}$ (bottom).
Left (or right) panels depict the BZ flux in the northern (or southern) hemispheres. 
Each curve denotes the BZ flux at colatitude $\theta$ as labeled.
}
    \label{fig:LBZ_Time_MA}
\end{figure*}

To smear out the variation, we take a moving average 
with a period of $5 GMc^{-3}$,
and plot the BZ fluxes in figure~\ref{fig:LBZ_Time_MA}.
In this particular figure, we compare the results
for three accretion rates:
the top, middle, and bottom panels show the BZ fluxes
at $\dot{m}=0.000250$, $0.000225$, and $0.000200$, respectively.
We find that the flux is enhanced for typically $140 \sim 180$ dynamical time scales,
and that the flux peaks in the middle latitudes during the enhancement
irrespective of the accretion rate,
as the blue dashed ($\theta=50^\circ$),
the blue solid ($\theta=60^\circ$), and 
the black dashed ($\theta=70^\circ$) curves indicate in each panel.
For example, at $\theta=60^\circ$, the FWHMs of the moving-averaged flux become
$157M$ and $149M$ in the northern and southern hemispheres, respectively,
when $\dot{m}=2.50 \times 10^{-4}$ (as the top two panels show).
They become $136M$ (northern) and $184M$ (southern) when $\dot{m}=2.25 \times 10^{-4}$
(middle two panels),
and $141M$ (northern) when $\dot{m}= 2.00 \times 10^{-4}$ (bottom left panel).
However, it is still not possible to measure the FWHM for the sourthern hemisphere
when $\dot{m}=2.00\times 10^{-4}$ (bottom right panel).
The flux distributes more symmetrically between northern and southern hemispheres,
as the accretion rate increases.

\subsubsection{Particle distribution and currents}
\label{sec:ptcl}
\begin{figure*}[t!]
\vspace*{0.0truecm}
\includegraphics[width=\textwidth, angle=0]{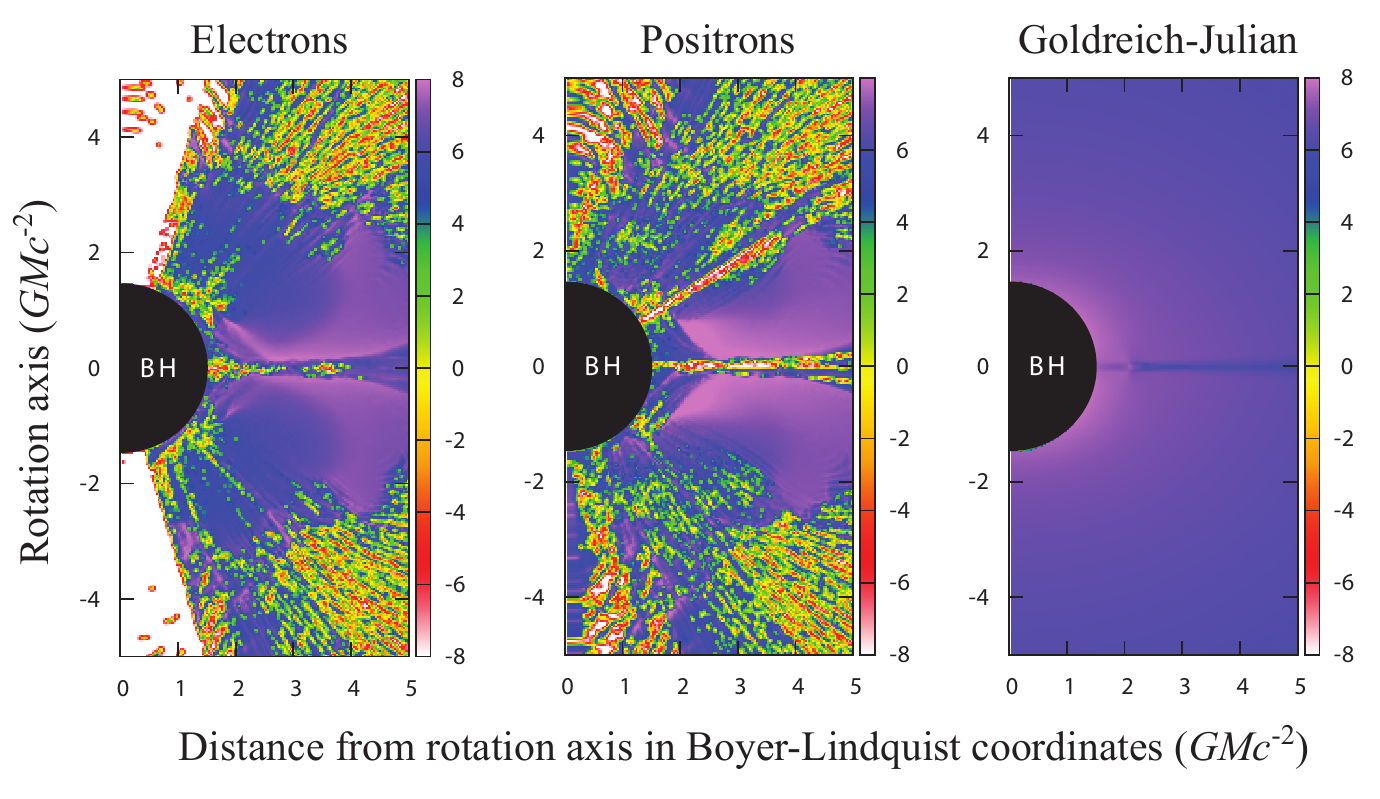}
\vspace*{0.0truecm}
\caption{
Densities of electrons (left) and positrons (middle), and 
the Goldreich-Julian density (right), 
$n_{\rm GJ}$, in ${\rm cm}^{-3}$ unit at $t=430M$.
Values are plotted in decadic logarithm.
The abscissa and ordinate are common with
figure~\ref{fig:Bp}.
}
    \label{fig:NeNp}
\end{figure*}

We next consider the distribution functions 
of $e^\pm$'s,
and the resultant current distribution.
Figure~\ref{fig:NeNp} shows 
the densities of electrons (left) 
and positrons (middle) at the burst peak, $t = 430M$,
in log scale. 
The right panel shows the GJ value at each point.
It follows that both electrons and positrons have
greater densities than the GJ value
particularly in the lower latitudes.
In the polar regions,
because of the polarization drift caused by
the varying meridional electric field
(and the constant radial magnetic field),
electrons migrate meridionally to accumulate in 
$\theta < 15^\circ$ and $\theta > 165^\circ$,
whereas positrons in
$15^\circ < \theta < 20^\circ$ and
$165^\circ > \theta > 160^\circ$ at $t \sim 430M$.
In the lower latitudes, the leptonic densities attain
$10^{7.7} \mbox{cm}^{-3}$.

\begin{figure*}[t!]
\vspace*{-0.0truecm}
\hspace{3.5cm}
\includegraphics[width=0.6\textwidth, angle=0]{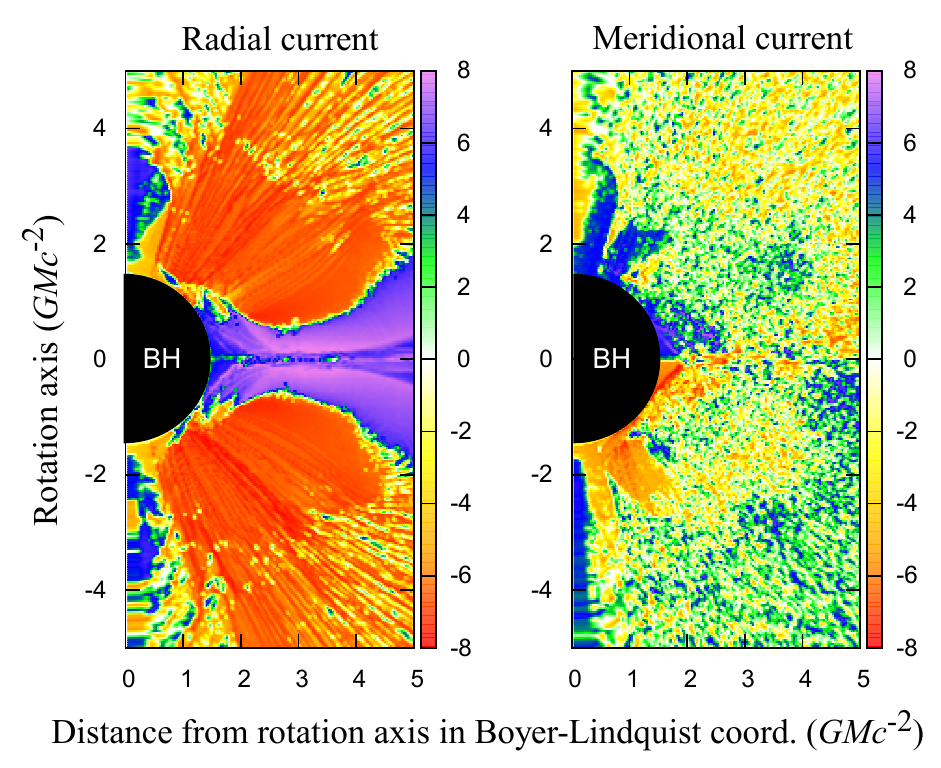}
\vspace*{-0.0truecm}
\caption{
Electric currents measured by ZAMO
in the poloidal plane at $t=430M$. 
The abscissa and ordinate are common with
figure~\ref{fig:Bp}.
The left panel shows the radial component,
$J^{\hat{r}}$, while 
the right one does the meridional component, $J^{\hat{\theta}}$.
To plot the values, we take the decadic lorarithm of 
the absolute value of a quantity, then put the same sign as
the quantity.
For example, in the left panel,
the value of $10^6$ (or $-10^6$) corresponds to 
an outward (or an inward) current density whose absolute value 
is $10^6 \mbox{ statampere cm}^{-2}$.
In the same way, in the the right panel, 
a positive (or a negative) $J^{\hat{\theta}}$
means an equator-ward current 
in the northern (or southern) hemisphere.
            }
    \label{fig:J1J2}
\end{figure*}

The charged leptons carry electric currents, 
as depicted in figure~\ref{fig:J1J2}.
In the left and right panels, 
we present the radial and meridional components
of the electric currents at each point in the ZAMO frame.
For a quantity $f(r,\theta)$, we plot
\begin{equation}
  F= \mbox{sign} \left\{ \lg \left[\max(|f|,1) \right] , f 
                 \right\},
  \label{eq:log_f}
\end{equation}
where $\mbox{sign}(a,b)= |a|$ if $b \ge 0$ and 
$= -|a|$ if $b<0$;
we set $f=J^{\hat{r}}$ and $J^{\hat\theta}$
for the left and right panels, respectively.
In the left panel,
the yellow-red regions show that currents flow inwards
in the middle latitudes,
while the blue-violet regions show
that they flow outwards in the lower latitudes.
These radial currents are closed by meridional currents
flowing within the ergosphere, as the right panel shows.
For example, in the right panel,
the blue (or red) region in the lower-middle
latitudes within $r < 2M$ in the northern (or southern)
hemisphere shows equator-ward meridional currents.
Because of this current closure,
it is confirmed that the BZ process is, indeed, facilitated.

\begin{figure*}[t!]
\vspace*{-0.0truecm}
\hspace{2.7cm}
\includegraphics[width=0.6\textwidth, angle=0]{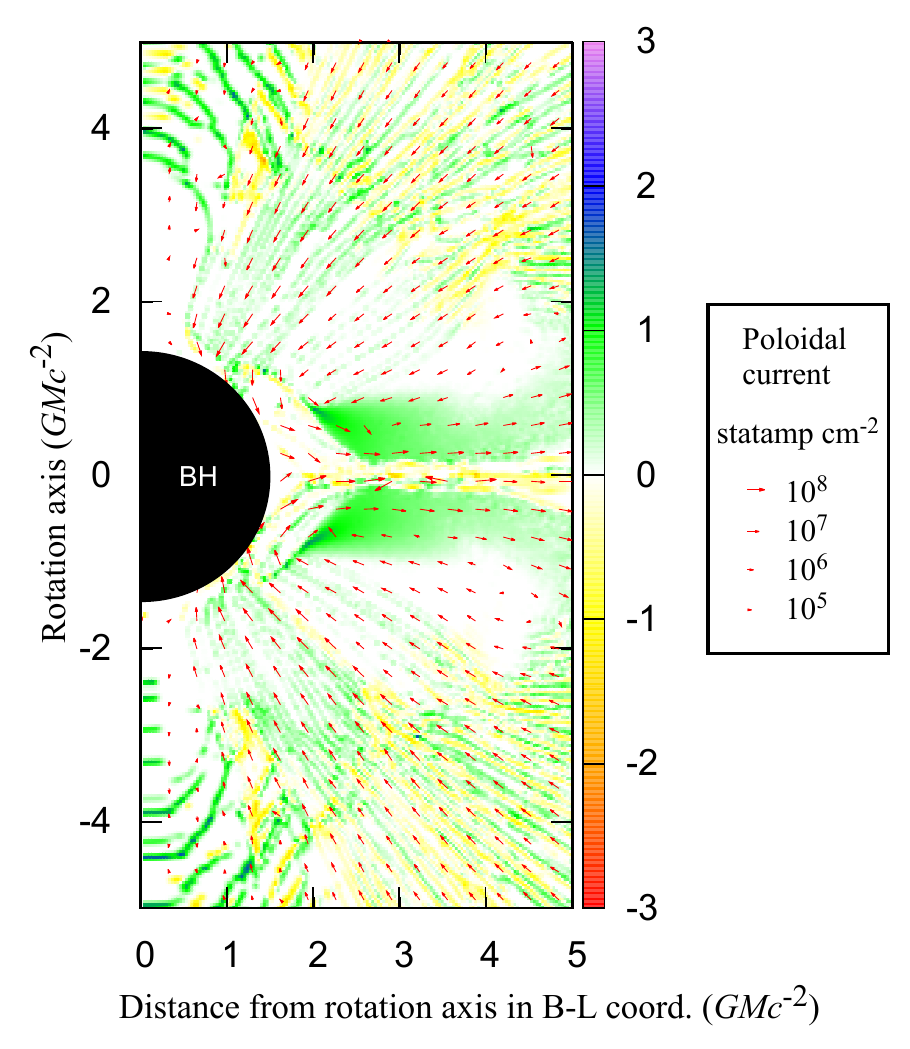}
\caption{
Dimensionless charge denisity $(n_+ - n_-)/n_{\rm GJ}$ (color image)
and poloidal electric current (red arrows).
The abscissa and ordinate are common with
figure~\ref{fig:Bp}.
The charge density is plotted in linear scale.
The green (or yellow) regions show positive (or negative) 
dimensionless charge densities.
The right panel shows four example arrow lengths
corresponding to the indicated strengths of 
the poloidal current densities
in $\mbox{statampere cm}^{-2}$.
            }
    \label{fig:chJp}
\end{figure*}

To grasp the current distribution more easily,
we plot the direction and strength of 
the poloidal currents as red arrows
in figure~\ref{fig:chJp}.
The current density is averaged over the area
in which we compute the direction and length of each arrow.
In the figure, the length of the arrows indicates the strength of
the current density in logarithmic scale,
as indicated by the right panel.
We can confirm the current pattern discussed 
in the foregoing paragraph.
It also follows that the averaged currents mostly flow 
in the middle and lower latitudes;
thus, the low density regions in the polar funnels 
do not essentially affect the entire structure of the magnetosphere.
In this figure, we also plot the charge density,
$(n_+ -n_-)/n_{\rm GJ}$, in color,
where $n_+$ and $n_-$ refer to the 
positronic and electronic number densities.
Values are plotted using the same method as figure~\ref{fig:J1J2}
(eq.~[\ref{eq:log_f}]).
It follows that the real charge density becomes even greater than the GJ
value (right panel of fig.~\ref{fig:NeNp}),
which indicates that the electron-positron plasmas 
become highly non-neutral near the BH.

\begin{figure*}[t]
\hspace{0.7cm}
\includegraphics[width=0.90\textwidth, angle=0]{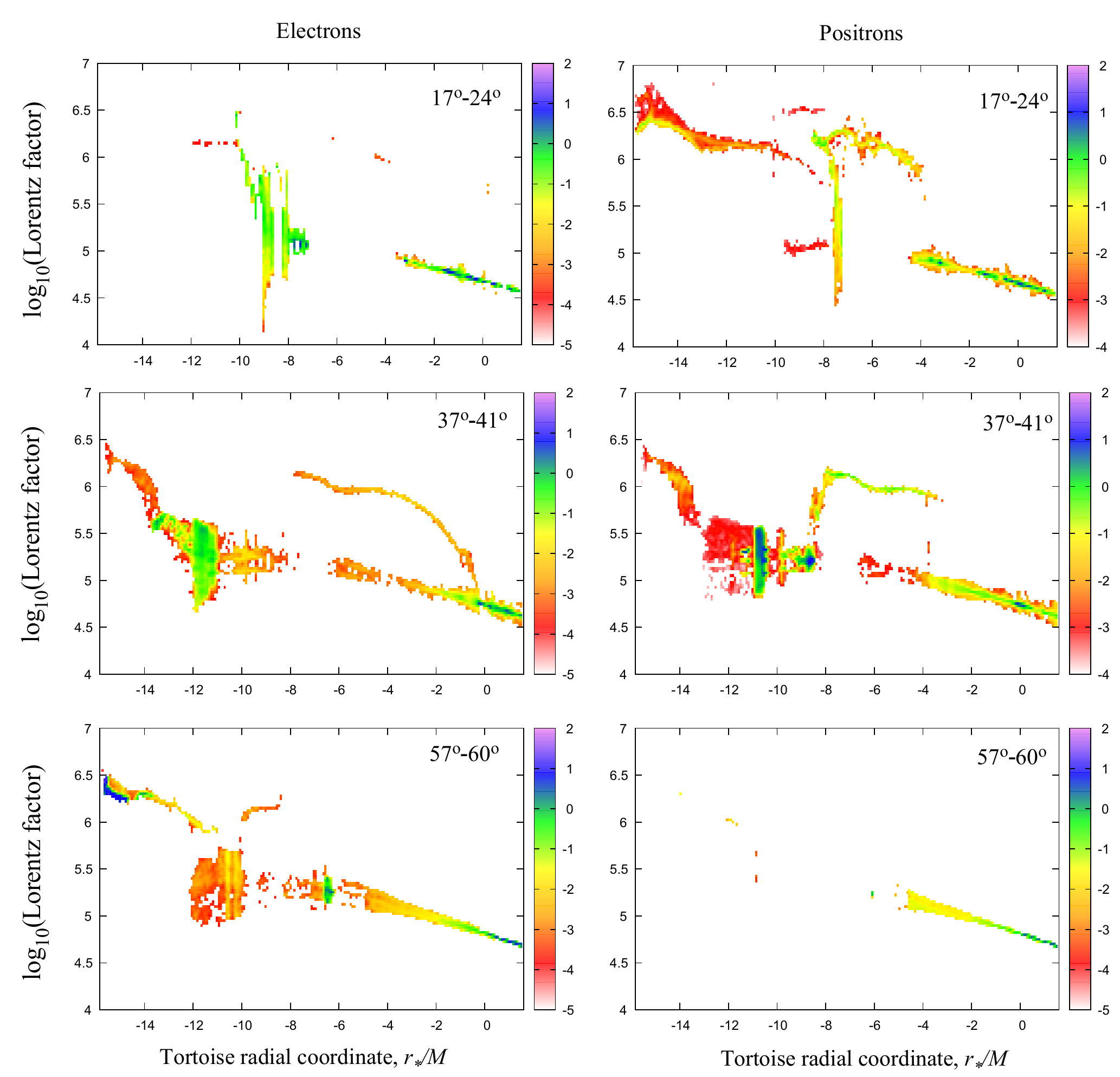}
\vspace*{-0.0truecm}
\caption{
Lorentz factor dependence of the distribution of electrons (left)
and positrons (right)
as a function of the dimensionless tortoise coordinate, $r_\ast / M$,
at elapse time $t=430 M$.
The color images show the logarithmic value of
the dimensionless distribution functions (see text)
along discrete colatitudes.
The top, middle, and bottom rows show the distribution function
between the meridional range
$16.606^\circ - 23.568^\circ$,
$37.678^\circ - 41.432^\circ$, and
$57.235^\circ - 60.034^\circ$, respectively.
            }
    \label{fig:Distr_LF}
\end{figure*}

We next consider the distribution functions 
of the charged leptons.
The dimensionless distribution functions
of $e^\pm$'s, $n_\pm$, are sliced between the colatitude 
$\theta_1 < \theta < \theta_2$,
\begin{equation}
  N_\pm (r_\ast,\gamma)
  \equiv
  \frac{1}{n_{\rm GJ}}
  \int_{\theta_1}^{\theta_2} d\theta 
    \frac{\partial^2 n_\pm (r_\ast,\theta,\gamma)}
         {\partial\theta \partial\gamma}.
  \label{eq:N_pm}
\end{equation}
In figure~\ref{fig:Distr_LF},
we present $N_-$ and $N_+$
as a function of $r_\ast$
and $\gamma$ (i.e., the Lorentz factor)
in the left and right columns, respectively.
The range of $\theta \in [\theta_1,\theta_2)$
increases from upper to lower rows
as described in the caption.
It shows that the Lorentz factors
are saturated at a terminal value at each point.
The terminal value is determined by the balance
between the electrostatic acceleration
and the synchro-curvature radiation drag force.
Since 
$\vert \mbox{\boldmath$E$} \cdot \mbox{\boldmath$B$}
 \vert$
increases with decreasing radius
significantly at $r_\ast < 0$
(or equivalently, at $r < 4.1M$),
particles gain large kinetic energies
inside $r < 4M$.
Nevertheless, by virtue of the residual 
magnetic-field-aligned electric field,
particle motion is kept relativistic in the entire region.
Lorentz factors attain $\gamma > 10^{4.5}$ 
in $r_\ast < 1M$ (or equivalently, in $r<4.7M$).

\begin{figure*}[t]
\hspace{0.7cm}
\includegraphics[width=0.9\textwidth, angle=0]{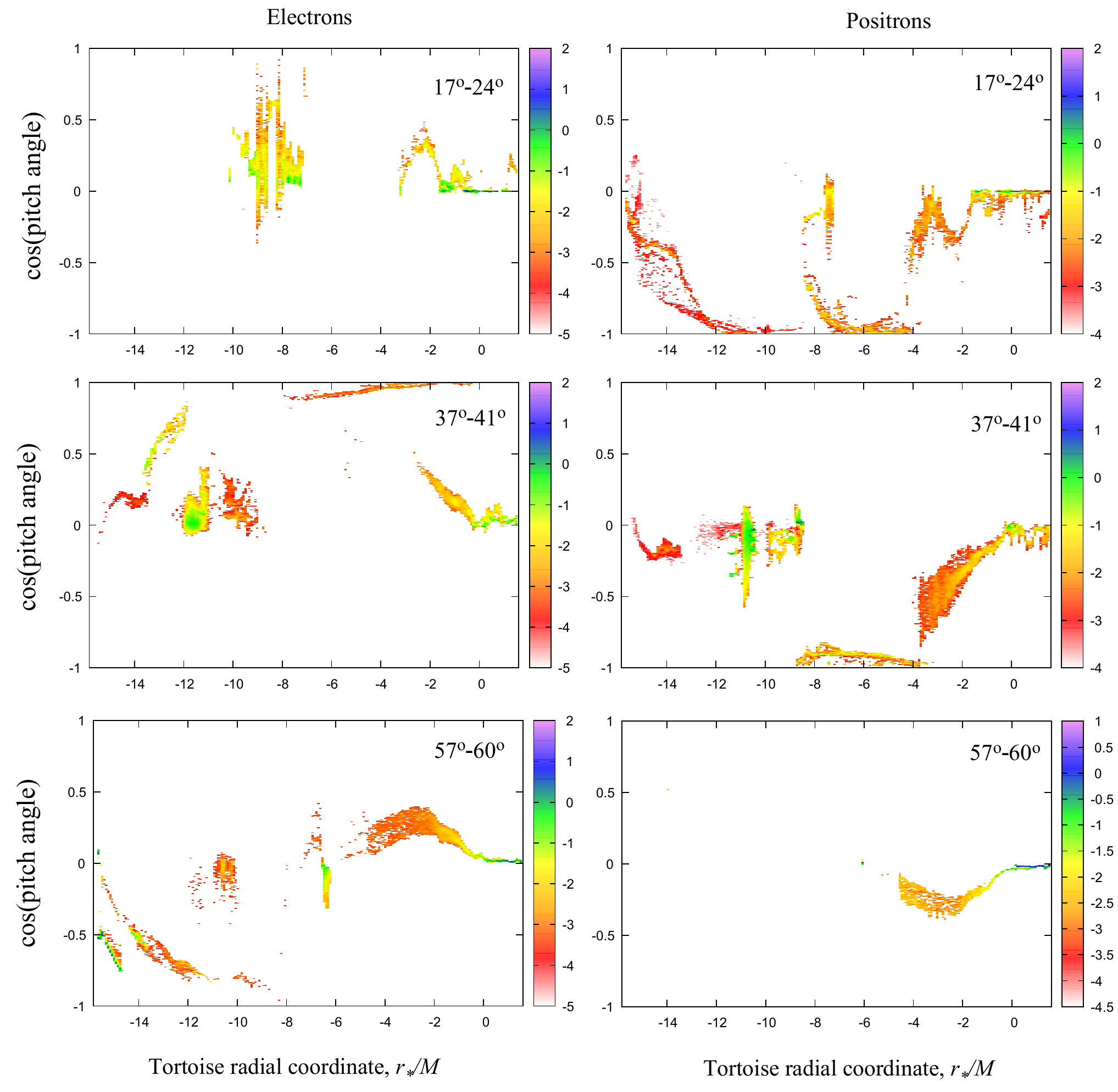}
\vspace*{-.0truecm}
\caption{
Similar figure as figure~\ref{fig:Distr_LF},
but the pitch-angle dependence of the distribution functions
of electrons (left) and positrons (right) at $t=430 M$.
The color images show the logarithmic value of
the dimensionless distribution functions
within the same meridional range as fig.~\ref{fig:Distr_LF}.
            }
    \label{fig:Distr_PA}
\end{figure*}

To examine the pitch-angle dependence,
in figure~\ref{fig:Distr_PA},
we plot $N_\pm(r_\ast,\chi)$,
where the pitch angle $\chi$ becomes $+1$ (or $-1$)
when a particle is moving outwards (or inwards) 
with a very small trans-magnetic-field momentum,
and becomes $0$ when it has no longitudinal momentum.
At $r_\ast > 0$,
particles have small trans-magnetic-field momenta.
However, in the higher-middle latitudes, $\theta \sim 40^\circ$,
the middle two panels show that
electrons (or positrons) migrate outwards with positive $\cos\chi$ 
(or inwards with negative $\cos\chi$),
indicating an inward poloidal current there.
Figures~\ref{fig:Distr_LF} and \ref{fig:Distr_PA} show
that particles are highly relativistic
with non-thermal and anisotropic distribution functions.

\section{Discussion}
\label{sec:disc}
To sum up,
we simulated the evolution of a BH magnetosphere
by a PIC scheme,
when a poloidal magnetic field is sustained 
by a disk toroidal current,
and when the electron-positron pair plasmas are steadily
supplied homogeneously per invariant volume basis.
Provided that the mass accretion rate is much less
than the Eddington rate,
both the electromagnetic fields
and the particle distribution functions exhibit rapid variability.
The rotational energy of the BH is, indeed, extracted
via the Blandford-Znajek process,
whose energy flux concentrates in the middle latitudes,
particularly during the flux-enhancement phase that lasts approximately
$160 \pm 20$ dynamical time scales.
We have demonstrated 
that the collision time scale is much longer than the
gyration time scale 
(i.e., the Ohm's law cannot be justified),
that the pair plasma is highly non-neutral,
that the particles' energy distribution is non-Maxwellian,
and that the momentum distribution is anisotropic.
Thus, we must discard the MHD approximation
when we consider the jet launching region 
around the BH whose mass accretion rate
is highly sub-Eddington.

In this section, 
we discuss the dominant radiative process 
in \S~\ref{sec:disc_radiation},
the validity of gridding in
\S~\ref{sec:disc_resolution}, 
comparison with other works in
\S~\ref{sec:disc_cf}, and
an implication to the collimation of VLBI jets
in \S~\ref{sec:disc_SMBH}.

\subsection{Dominant radiative process}
\label{sec:disc_radiation}
Although the synchro-curvature radiation process is incorporated in 
the radiative reaction force, $F_{\rm rad}^j$,
ICS are not considered as a radiation drag.
Thus, we have to confirm that the ICS process
is negligible compared with the synchro-curvature process
around stellar-mass BHs accreting at $\dot{m} \ll 1$.
Here we briefly compare the pure curvature,
pure synchrotron, and ICS processes,
and discuss the dominant process.
To make a general discussion,
we adopt the actual electron mass $m_{\rm e}=m_{\rm p}/1836$,
instead of $m_{\rm p}/20$, in this subsection.

The magnitude of the radiation drag force 
due to the pure curvature process is given by
\begin{eqnarray}
  F_{\rm curv}
  &=& \frac23 e^2 \frac{\gamma^4}{\rho_{\rm c}^2}
  \nonumber\\
  &=& 1.70 \times 10^{-4}
      \left(\frac{\rho_{\rm c}}{2M}\right)^{-2}
      M_1{}^{-2} \gamma_7{}^4 \mbox{ dyn}, \qquad
  \label{eq:F_curv}
\end{eqnarray}
where $\rho_{\rm c}$ refers to the 
curvature radius of the particle's center of gyration,
$2M=2MGc^{-2}$ the Schwarzschild radius,
and $\gamma_n \equiv \gamma/10^n$.
If this force balances with the electrostatic force,
\begin{equation}
  e E_\parallel
  = 4.80 \times 10^{-5}
    \vert E_{\parallel,5} \vert
    \mbox{ dyn},
  \label{eq:electrostatic}
\end{equation}
we find that the particles saturate 
at the Lorentz factor,
\begin{equation}
  \gamma_{\rm curv}
  = 7.28 \times 10^6
    \left(\frac{\rho_{\rm c}}{2M}\right)^{1/2}
    M_1{}^{1/2}
    E_{\parallel,5}{}^{1/4},
    \label{eq:gamma_curv}
\end{equation}
where 
$E_{\parallel,5} \equiv E_\parallel 
  / (10^5 \mbox{ statvolt cm}^{-1}) $.

The magnitude of the radiation-drag force
due to the pure synchrotron process is given by
\begin{eqnarray}
  F_{\rm sync}
  &=& \frac23 r_0{}^2 \gamma^2 B^2 \sin^2\chi
  \nonumber\\
  &=& 5.29 \times 10^{-5}
    \gamma_5{}^2 B_6{}^2
    \frac{\sin^2 \chi}{0.1}
    \mbox{ dyn},
  \label{eq:F_sync}
\end{eqnarray}
where $r_0$ denotes the classical electron radius,
and $\chi$ the pitch angle.
Equating equations~(\ref{eq:electrostatic})
and (\ref{eq:F_sync}), we obtain
the terminal Lorentz factor
\begin{equation}
  \gamma_{\rm sync}
  = 9.52 \times 10^4 B_6{}^{-1}
    \left( \frac{\sin^2\chi}{0.1} \right)^{-1/2}
    E_{\parallel,5}^{1/2}.
  \label{eq:gamma_sync}
\end{equation}
If $\gamma_{\rm sync} < \gamma_{\rm curv}$,
the pure synchrotron process dominates 
the pure curvature process.
If $m_{\rm e}=m_{\rm p}/1836$,
We find that the condition
$\gamma_{\rm sync} < \gamma_{\rm curv}$
is satisfied,
because 
\begin{equation}
  B_6 \left( \frac{\sin\chi}{0.1} \right)
  E_{\parallel,5}^{-1/4}
  \left( \frac{\rho_{\rm c}}{2M} \right)^{1/2}
  > 1
  \label{eq:sync_dom}
\end{equation}
is satisfied by the vast majority of the particles.
In another word, the synchro-curvature process
reduces to the pure synchrotron process
for most of the particles.
However, since we assume heavy electrons in this paper,
the increased gyro radius makes the synchrotron process less efficient;
thus, the radiation drag force is given by the synchro-curvature process
in general for heavy electrons. 

We next compare the pure synchrotron process
with the ICS.
At radius $r$ from the central BH,
the number density of the ADAF synchrotron photons
are given by 
\citep{Mahadevan:1997:ApJ}
\begin{equation}
  N_{\rm ph}
  = \frac{L_{\rm sync}}
         {4\pi r^2 h \nu c},
  \label{eq:Nph}
\end{equation}
where $h \nu$ refers to the energy of the
photons emitted from the ADAF via the synchrotron
process.
The synchrotron luminosity is given by
\begin{equation}
  L_{\rm sync}
  = 1.67 \times 10^{36}
    M_1{}^{1/2} \dot{m}^{3/2}
    \mbox{ ergs s}^{-1}.
  \label{eq:Lsync}
\end{equation}
The typical photon energy is given 
by equation~(22) of \citet{Mahadevan:1997:ApJ}.
Accordingly, we obtain
\begin{equation}
  N_{\rm ph}
  = 1.55 \times 10^{20}
    T_{\rm e,9}{}^5 M_1 \dot{m}
  \mbox{ photons cm}^{-3},
  \label{eq:Nph_2}
\end{equation}
where $T_{\rm e,9}$ refers to
the electron temperature, $T_{\rm e}$,
normalized by $10^9 K$.
The ICS drag force per electron (or positron)
is given by
\begin{equation}
  F_{\rm ICS}
  \approx N_{\rm ph} \sigma_{\rm KN} 
          \gamma m_{\rm e} c^2,
  \label{eq:F_ICS}
\end{equation}
where 
\begin{equation}
  \sigma_{\rm KN}
  \approx \frac38 \sigma_{\rm T}
          x^{-1} (\ln 2x + \frac12 ),
  \label{eq:KN}
\end{equation}
refers to the Klein-Nishina cross section, and
$x \approx \gamma$ holds on average.
We thus have
\begin{equation}
  F_{\rm ICS}
  \approx 3.89 \times 10^{-10} 
    M_1 \dot{m} T_{\rm e,9}{}^5
  \label{eq:F_ICS_2}
\end{equation}
We thus obtain
\begin{equation}
  \gamma_5{}^2
  B_6{}^2
  \left( \frac{\sin\chi}{0.1} \right)^2 
  M_1^{-1}
  \dot{m}^{-1}
  T_{\rm e,9}{}^{-5}
  \gg 10^{-5}
  \label{eq:ICSvsS}
\end{equation}
On these grounds, 
we obtain $F_{\rm sync} \gg F_{\rm curv} \gg F_{\rm ICS}$
for the actual electron mass.
Note that the Lorentz factor will increase
with decreasing electron mass.
However, for heavy electrons,
energy transfer efficiency increases due to their greater mass ratio to protons.
Thus, ICS may not be negligible even for stellar-mass BHs.
However, we neglected ICS in this paper, considering future extension 
to smaller electron masses.

In short, for the actual electron mass, 
it is possible that the ICS process is negligible
compared to the synchro-curvature process,
when we consider stellar-mass BHs in a quiescent state.
However, for supermssive BHs,
we generally obtain 
$F_{\rm ICS} > F_{\rm curv} \gg F_{\rm sync}$
because of the large curvature radius
and the weak magnetic field strength
\citep{Hirotani:2016:ApJ} for the actual electron mass.

\subsection{Grid interval versus skin depth}
\label{sec:disc_resolution}
Let us compare the invariant grid interval
$ \epsilon r_{\rm g} \equiv
  \Delta_s
  \equiv
  {\rm max} \left( \sqrt{g_{rr}} \Delta_r, 
                   \sqrt{g_{\theta\theta}} \Delta_\theta
            \right)
$
with the skin depth, $l_{\rm p}$ 
(eq.~\ref{eq:skin_depth}),
where $\Delta_r$ and $\Delta_\theta$ denote 
the interval in 
$r$ and $\theta$ coordinates, respectively.
Representative values of the dimensionless function $\epsilon(r,\theta)$ are
presented in table~1. 
Because of a constant $\sin\theta \Delta_\theta$ gridding, 
$\Delta_\theta$ increases with decreasing $\sin\theta$. 
Note that the dimensionless grid interval $\epsilon$ is defined 
independently from the BH mass.

Adopting heavy electrons, $m_{\rm e}= m_{\rm p}/20$,
and normalizing the skin depth with $r_{\rm g}$, we obtain
\begin{equation}
  \frac{l_{\rm p}}{r_{\rm g}}
  = 3.4 \left( \gamma_5 n_5{}^{-1} \right)^{1/2}
  \label{skin_depth_2}
\end{equation}
where $\gamma_\nu \equiv \langle\gamma\rangle / 10^\nu$ and
$n_\nu \equiv n_\pm / 10^\nu $;
$\langle\gamma\rangle$ denotes the averaged Lorentz factor.
The highest density region appears in the lower latitudes
at $r<4M$, i.e., at $r_\ast < -0.25M$.
In this region, $n < 10^{7.7} \mbox{ cm}^{-3}$
and $\langle \gamma \rangle > 10^{4.7}$;
thus, we obtain
$l_{\rm p} > 0.108 r_{\rm g} > 5 \Delta_{\rm s}$
there.
Other regions have greater skin depth.
Accordingly, the skin depth can be resolved
at every point (and in fact at every time step)
during the PIC simulation,
although it is marginal when $l_{\rm p} \sim 5 \Delta_{\rm s}$ happens
at the flux peak.

\begin{center}
\vspace*{1.0truecm}
{\bf Table~1}\\
Invariant grid interval, 
$\epsilon(r,\theta)=\Delta_s / r_{\rm g}$
\begin{tabular}{ccccc}\hline\hline
  $r/r_{\rm g}$ & $\theta=2.62^\circ$  & $29.0^\circ$
                       & $60.0^\circ$  & $90.0^\circ$
  \\ \hline
  8.507  & .2762    & .0262    & .0261   & .0260  \\
  4.099  & .1355    & .0217    & .0215   & .0214  \\
  2.021  & .0714    & .0122    & .0117   & .0114  \\
  1.471  & .0557    & .0036    & .0028   & .0026  \\
  \hline
  \label{tbl:grid_interval}
\end{tabular}
\end{center}

\subsection{Comparison with previous works}
\label{sec:disc_cf}
Let us compare the present work with
recent two works on 2-D GR PIC simulations
\citep{Parfrey:2019:PhRvL,Crinquand:2020:PhRvL}.

First, we discuss the difference with
\citet{Parfrey:2019:PhRvL}.
Ignoring radiative transfer, 
and considering instead an injection of pairs
whose rate is proportional
to the local magnetic-field-aligned electric field,
they performed 2-D GR PIC simulations 
of the magnetosheres of an extremely rotating
BH with $a=0.999M$.
The BH mass is not explicitly specified in their paper,
but presumably supermassive.
They adopted an extremely small magnetic field strength
to emphasize the Penrose process,
an energy-extraction mechanism from rotating BHs
when particles fall onto the horizon 
with negative energies measured at infinity.
Particles were created in the reconnecting current sheet
on the equatorial plane.
The magnetization parameter was $2000$;
that is, the magnetosphere was still magnetically dominated.
Magnetic field configuration is initially the Wald's vacuum solution,
which is produced by a toroidal ring current flowing on the equatorial
plane at large distances.
Then the field lines are bend back toward the BH to penetrate the horizon.
For M87*, their dimensionless magnetic field strength,
$\tilde{B}_0 = 10^3$ corresponds to the actual strength
$B= \tilde{B}_0 (m_{\rm e}c^2/e r_{\rm g})
 \sim 10^{-9}$~G, which is about $10^{-11}$ times
smaller than what is observed 
\citep{EHT:2019:ApJL}.
Under this condition,
they found that the charged leptons created in the current sheet
plunge onto the horizon with negative energies
as a result of the interaction with the electromagnetic field,
and that the Penrose process contribute
in the extraction of the energy and angular momentum 
from a maximally rotating BH.

In the present paper, on the other hand,
we consider a stellar-mass BH.
In this case, we can resolve the skin depth
with a realistic magnetic field strength,
$B= B_{\rm eq}$.
Accordingly, 
the magnetization parameter becomes of the order of $10^7$;
thus, we did not mention about the particle contribution 
on the energy extraction from a rotating BH.
Nevertheless, particle energy flux does become negative
in our simulation as well;
thus, the Penrose process is indeed working, 
although their contribution is negligible.
We assume a homogeneous and constant plasma supply in the present work, 
which forms a constast to \citet{Parfrey:2019:PhRvL},
who considered pair creation in the reconnecting current sheet.
The magnetic field is assumed to be fixed on the poloidal plane
during the simulation in the present paper.

Second, let us discuss the difference with 
\citet{Crinquand:2020:PhRvL}.
In their 2D GR PIC code, 
implementing inverse-Compton scatterings (ICS)
and photon-photon pair production self-consistently,
they solved the PIC equations together with
the radiative transfer equation.
Electron-positron pair plasmas are supplied in response to gap opening.
They presumed supermassive BHs.
In this case, ICS dominates the pure-curvature radiation
\citep{Hirotani:2016:ApJ},
where the particle's pitch angles become small enough
due to the ICS and continuous acceleration along the
magnetic field lines \citep{bes92}.
Thus, their treatment, which takes account of 
only the ICS as radiative processes, can be justified.
\citet{Crinquand:2020:PhRvL} assumed 
a true monopole magnetic field,
whose radial component point outwards 
in the entire magnetosphre.
Accordingly, there appear no current sheets in their
solution.
Electric currents flow inwards (or outwards) 
in the northern (or southern) hemisphere, 
leading to a negative 
$F^{r \theta}$ in the entire magnetosphere.
Since $E_\theta=F_{\theta t}>0$ holds in both hemispheres,
the resultant Poynting flux is also positive
in both hemispheres.
In their Zelton code, they solved all the six components of 
the electromagetic fields,
computing the evolution of $F_{\varphi t}$ from the toroidal component
of the current density $J^\varphi$ that is constructed from the 
azimuthal motion of the charged particles.

In the present paper, on the other hand,
instead of solving the radiative transfer equation,
we simply adopt a uniform and constant
pair production in the particle source term.
We consider a stellar-mass BH,
in which case synchro-curvature process
becomes non-negligible compared to ICS 
\citep{Hirotani:2016:ApJ}.
We assume a split-monopole magnetic field,
whose radial component is positive (or negative)
in the northern (or southern) hemisphere.
Accordingly, there appears a current sheet on the equator.
Electric currents flow inwards in the higher latitudes
in both hemispheres, 
and flow outwards in the lower latitudes,
leading to a negative (or positive) $F^{r\theta}$
in the northern (or southern) hemisphere,
and a positive BZ flux in both hemispheres,
as discussed in \S\ref{sec:acc}.

Let us briefly point out the limitation of the present analysis.
Unlike the Zelton code, 
we solved only three of the six components
of the electromagnetic fields in the present paper, putting
$F_{\varphi t}=0$, and hence 
$\partial_t F_{\theta \varphi}= \partial_t F_{\varphi r}= 0$ 
throughout the simulation, as the first step.
If the local physics (as supposed in recent 1-D GR PIC simulations) does not significantly affect the entire structure
of the magnetosphere, 
such an assumption of a fixed poloidal magnetic field could be justified.
However, when we apply the present 2D method to a global magnetosphere,
toroidal currents carried by the charged leptons at each position
can alter the entire magnetic field configuration on the poloidal plane.
Thus, in the same manner as in recent 1-D models
\cite{Levinson:2018:AA,Chen:2020:ApJ,Kisaka:2020:arXiv}, 
we must draw attention to the limitation of $F_{\varphi t}=0$.
In the next paper, we will extend our analysis to 
the case $F_{\varphi t} \ne 0$,
constructing $J^\varphi$ at each position from the actual motion of
the charged leptons, and
solving all the six components of the electromagnetic fields.

\subsection{Implication for supermassive black holes}
\label{sec:disc_SMBH}
Let us finally discuss what can be expected 
if the present results obtained for stellar-mass BHs
can be applied to supermassive BHs.
We have demonstrated that the BH's rotational energy is efficiently 
extracted along the magnetic field lines that cross the event horizon 
in the middle latitudes.
Let us discuss an implication of this result on the formation of 
limb-brightened jets.
It has been revealed that the innermost region of the M87 jet
exhibits a limb-brightened structure
by the VLBA observations in 15--43~GHz 
\citep{Kovalev:2007:ApJL,Ly:2007:ApJ,Junor:1999:Natur,
       hada11,hada13,Walker:2018:ApJ}.
At 86~GHz,
this limb-brightened structure is already well developed
at 0.15~mas from the VLBI core,
where the corresponding apparent opening angle 
becomes approximately $100^\circ$ \citep{hada16}.
If a relatively large viewing angle of 
$\theta_{\rm view} \sim 30^\circ$ is adopted 
\citep{Ly:2007:ApJ,hada16},
the jet has a deprojected opening angle $\chi_{\rm open} \sim 50^\circ$
at the deprojected distance $z=84 GMc^{-2}$.
However, if a smaller view angle, $\theta_{\rm view} \sim 17^\circ$,
is adopted 
\citep{Biretta:1999:ApJ,Wang:2009:MNRAS,Perlman:2011:ApJ,
       Nakamura:2014:ApJ,Mertens:2016:AA,Walker:2018:ApJ},
we obtain $\chi_{\rm open} \sim 30^\circ$ 
at $z=108 GMc^{-2}$.

\begin{figure*}
\hspace{4.3cm}
\includegraphics[width=1.0\columnwidth, angle=0]{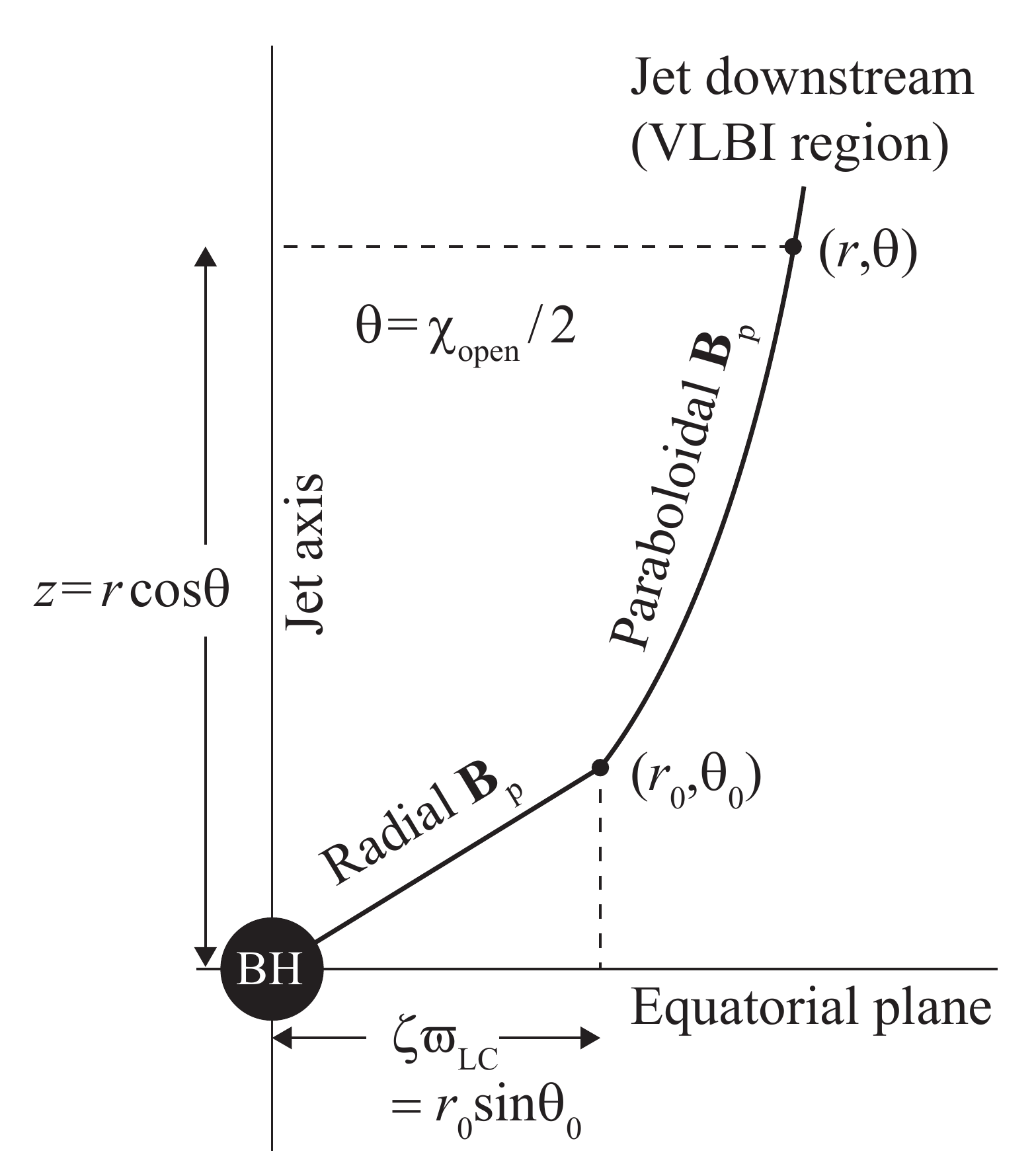}
\caption{
Schematic picture of the jet downstream region.
The jet shape is assumed to be radial 
inside the radius $r_0$,
and paraboidal outside of it.
The jet is observed with VLBI
at radius $r$.
Both $\theta$ and $\theta_0$ are measured
from the jet axis,
which coincides with the
BH's rotation axis.
            }
    \label{fig:jet}
\end{figure*}

If a jet begins to collimate outside the outer light surface
\citep{came86b},
we may assume that the jet is radial within the distance
$\zeta \varpi_{\rm LC}$ from the rotation axis,
and becomes paraboloidal outside of it
\citep{Asada:2012:ApJL,hada13,
       Nakamura:2013:ApJ,Asada:2016:ApJ,Nakamura:2018:ApJ},
where $\varpi_{\rm LC} \equiv c / \Omega_{\rm F}$
denotes the typical radius of the outer light surface.
Figure~\ref{fig:jet} sketches the geometry
of this jet downstream region.
Assuming that we observe the jet at the position
($r$,$\theta$),
we can express $\zeta$ in terms of
the observables $z$ and $\theta=\chi_{\rm open}/2$
as
$r(1-\cos\theta)= 
 (\zeta \varpi_{\rm LC}/\sin\theta_0)(1-\cos\theta_0)$,
which gives
\begin{equation}
  \zeta 
  \approx 
  \frac{z}{\varpi_{\rm LC}}
  \frac{1-\cos\theta}{\cos\theta}
  \frac{\sin\theta_0}{1-\cos\theta_0},
  \label{eq:jet}
\end{equation}
where $z=r\cos\theta$.
If $\Omega_{\rm F} \approx 0.5 \omega_{\rm H}$, 
$a \approx 0.9M$ gives
$\varpi_{\rm LC} \approx 6.4 GM/c^2$.
(Note that $\Omega_{\rm F}=F_{t\theta}/F_{\theta\varphi}$ is not assumed
 but solved in the PIC simulation.) 
If the magnetic field line with the footpoint angle
$\theta_0 \approx 60^\circ$ (or $75^\circ$)
is brightened at downstream $z$ with half opening angle
$\theta= \chi_{\rm open}/2$, 
we obtain
$\zeta \approx 2.4$ 
(or $\zeta \approx 1.8$)
for $\theta_{\rm view}=30^\circ$,
and 
$\zeta \approx 1.0$
(or $\zeta \approx 0.78$) 
for $\theta_{\rm view}=17^\circ$,
using the 86~GHz VLBI observations.
On these grounds,
if the BZ flux concentrates in the middle
latitudes also around supermassive BHs,
it is possible that the M87 jet begins to collimate
slightly outside the outer light surface,
typically within the distance $2.4 \varpi_{\rm LC}$ 
from the rotation axis,
which may become a good target of the Event Horizon Telescope
and GRAVITY.

\acknowledgments

We thank the anonymous referee for careful reading and valuable comments to this manuscript. The authors acknowledge grant support to the CompAS/Antares group, and the access to high-performance computing facility from the Institute of Astronomy and Astrophysics in Academia Sinica (ASIAA). The authors acknowledge grant support from Ministry of Science and Technology (MoST) under 105-2119-M-001-044-MY3, 108-2112-M-001-009- and 109-2112-M-001-028-.

\appendix

\section{ZAMO-measured quantities}
\label{sec:app_ZAMO}
We give the expressions of ZAMO-measured quantities, which are indicated 
by tilde ($\,\tilde{}\,$), in this appendix section.
Using equations~(\ref{eq:rotationg_e0z})--(\ref{eq:rotationg_e3z}), we obtain
\begin{equation}
  u_{\hat{r}}= \sqrt{\frac{\Delta}{\Sigma}} u_r,
  \label{eq:u_r}
\end{equation}
\begin{equation}
  u_{\hat{\theta}}= \sqrt{\frac{1}{\Sigma}} u_\theta,
  \label{eq:u_theta}
\end{equation}
\begin{equation}
  u_{\hat{\varphi}}= \sqrt{\frac{A}{\Sigma}} \sin\theta u_\varphi,
  \label{eq:u_phi}
\end{equation}
Or equivalently, their contravariant components become
\begin{equation}
  u_{\hat{r}} =
  u^{\hat{r}} \equiv \sqrt{g_{rr}} u^r,
  \label{eq:def_ur}
\end{equation}
\begin{equation}
  u_{\hat{\theta}}=
  u^{\hat{\theta}} \equiv \sqrt{g_{\theta\theta}} u^\theta,
  \label{eq:def_uth}
\end{equation}
\begin{equation}
  u_{\hat{\varphi}}=
  u^{\hat{\varphi}} 
  \equiv \sqrt{g_{\varphi\varphi}} 
         (u^\varphi -\omega u^t),
  \label{eq:def_uph}
\end{equation}
\begin{equation}
  u^j \equiv \gamma v^j,
  \label{eq:def_u}
\end{equation}
where
\begin{equation}
  \gamma \equiv \frac{1}{\sqrt{1-\vec{v}\cdot\vec{v}}}.
  \label{eq:def_Lf}
\end{equation}

In the same way, the tetrad transformation law gives
the following expressions of the elctromagnetic fields in ZAMO:
\begin{eqnarray}
  \tilde{E}_{\hat r} 
  &=&
  \mbox{\boldmath$F$}(\tilde{\mbox{\boldmath$e$}}_{(\hat{r})},
                      \tilde{\mbox{\boldmath$e$}}_{(\hat{t})})
  \nonumber\\
  &=&
  \frac{\sqrt{(r^2+a^2)^2 -\Delta a^2 \sin^2\theta}}
       {\Sigma}
  \left( F_{r t} + \omega F_{r \varphi} \right)
  \nonumber\\
  \label{eq:Er_ZAMO}
\end{eqnarray}
\begin{eqnarray}
  \tilde{E}_{\hat \theta} 
  &=&
  \mbox{\boldmath$F$}(\tilde{\mbox{\boldmath$e$}}_{(\hat{\theta})},
                      \tilde{\mbox{\boldmath$e$}}_{(\hat{t})})
  \nonumber\\
  &=&
  \frac{\sqrt{(r^2+a^2)^2 -\Delta a^2 \sin^2\theta}}
             {\sqrt{\Delta} \Sigma}
  \left( F_{\theta t} + \omega F_{\theta \varphi} \right)
  \nonumber\\
  \label{eq:Eth_ZAMO}
\end{eqnarray}
\begin{equation}
  \tilde{E}_{\hat \varphi} 
  = \mbox{\boldmath$F$}(\tilde{\mbox{\boldmath$e$}}_{(\hat{\varphi})},
                        \tilde{\mbox{\boldmath$e$}}_{(\hat{t})})
  = \frac{1}{\rho_{\rm w}} F_{\varphi t}
  = \frac{1}{\sqrt{\Delta} \sin\theta} F_{\varphi t}
  \label{eq:Eph_ZAMO}
\end{equation}
\begin{equation}
  \tilde{B}_{\hat r} 
  = {}^\ast \!
    \mbox{\boldmath$F$}(\tilde{\mbox{\boldmath$e$}}_{(\hat{t})},
                        \tilde{\mbox{\boldmath$e$}}_{(\hat{r})})
  = F_{\hat{\theta}\hat{\varphi}}
  = \frac{1}{\sqrt{\Sigma g_{\varphi\varphi}}} F_{\theta\varphi},
  \label{eq:Br_ZAMO}
\end{equation}
\begin{equation}
  \tilde{B}_{\hat \theta} 
  = {}^\ast \!
    \mbox{\boldmath$F$}(\tilde{\mbox{\boldmath$e$}}_{(\hat{t})},
                        \tilde{\mbox{\boldmath$e$}}_{(\hat{\theta})})
  = -F_{\hat{r}\hat{\varphi}}
  = \frac{\sqrt{\Delta}}{\sqrt{\Sigma g_{\varphi\varphi}}} F_{\varphi r},
  \label{eq:Bth_ZAMO}
\end{equation}
\begin{equation}
  \tilde{B}_{\hat \varphi} 
  = {}^\ast \!
    \mbox{\boldmath$F$}(\tilde{\mbox{\boldmath$e$}}_{(\hat{t})},
                        \tilde{\mbox{\boldmath$e$}}_{(\hat{\varphi})})
  = \sqrt{g_{rr} g_{\theta\theta}} F^{r \theta}
  = \frac{\Sigma}{\sqrt{\Delta}} F^{r \theta},
  \label{eq:Bph_ZAMO}
\end{equation}
where ${}^\ast \! \mbox{\boldmath$F$}$ refers to the Maxwell tensor,
which is the dual of the Faraday tensor $\mbox{\boldmath$F$}$.
For example, figure~\ref{fig:Bp} depicts the distribution of
equations~(\ref{eq:Br_ZAMO}) and (\ref{eq:Bth_ZAMO}) on the poloidal plane.
We could use equations~(\ref{eq:def_ur}) and (\ref{eq:def_uth}) to compute 
$J^{\hat{r}}$ and $J^{\hat{\theta}}$, which are depicted 
in figures~\ref{fig:J1J2}--\ref{fig:chJp}.
However, in actual calculations, we compute $J^r$ and $J^\theta$ from
equations~(\ref{eq:drdt}) and (\ref{eq:dthdt}) and convert them into 
the ZAMO-measured quantities,
using the transformation law of the one-form bases that are dual to 
the ZAMO's tetrad, (\ref{eq:rotationg_e0z})--(\ref{eq:rotationg_e3z}).

\section{Stationary, vacuum magnetosphere}
\label{sec:appendix_stationary}
In this appendix section, we formulate the basic equations
to solve the initial, stationary electromagnetic fields.

\subsection[]{The Gauss's law}
\label{sec:gauss}
The inhomogeneous part of the Maxwell equations 
can be written as
\begin{equation}
  \nabla_{\mu} F^{\nu\mu}=\frac{4\pi}{c} J^\nu,
  \label{eq:Poisson_1}
\end{equation}
where $\nabla_{\mu}$ denotes the covariant derivative with respect to
the coordinate variable $x^\mu$,
$F^{\nu\mu}$ the electro-magnetic field strength tensor,
and $J^\nu$ the four current density.
Putting $\nu=0$, we obtain the Gauss's law,
\begin{equation}
  \frac{1}{\sqrt{-g}} \partial_{\mu} (\sqrt{-g} F^{0 \mu})
  =4\pi \rho,
  \label{eq:Poisson_2}
\end{equation}
where $\rho$ refers to the electric charge density, and
$\sqrt{-g}= \Sigma \sin\theta$.
In the Kerr spacetime, we obtain
\begin{eqnarray}
  F^{0 1}
  &=&  
  F^{t r} = \frac{g^{rr}}{\rho_{\rm w}{}^2} 
    \left( -g_{\varphi\varphi} F_{t r} + g_{t \varphi} F_{\varphi r} \right),
  \label{eq:EM_1}
  \\
  F^{0 2}
  &=& 
  F^{t \theta}  = \frac{g^{\theta\theta}}{\rho_{\rm w}{}^2} 
    \left( -g_{\varphi\varphi} F_{t \theta} 
           +g_{t \varphi} F_{\varphi \theta} \right),
  \label{eq:EM_2}
  \\
  F^{0 3}
  &=& 
  F^{t \varphi} = \frac{1}{\rho_{\rm w}{}^2} F_{\varphi t}
  \label{eq:EM_3}
\end{eqnarray}

At $t=0$, 
we assume stationary and axisymmetric magnetospehre, 
$\partial_t= \partial_\varphi=0$, to obtain
\begin{eqnarray}
  F_{r t} 
  &=&  
  A_{t,r}-A_{r,t} = A_{t,r} = \partial_r A_t,
  \label{eq:EM_4}
  \\
  F_{\theta t} 
  &=&  
  A_{t,\theta}-A_{\theta,t} = A_{t,\theta} = \partial_\theta A_t,
  \label{eq:EM_5}
  \\
  F_{\varphi t} 
  &=&  
  A_{t,\varphi}-A_{\varphi,t}=0,
  \label{eq:EM_6}
  \\
  F_{\varphi r} 
  &=&  
  A_{r,\varphi}-A_{\varphi,r} = -\partial_r A_\varphi,
  \label{eq:EM_7}
  \\
  F_{\theta \varphi} 
  &=&  
  A_{\varphi,\theta}-A_{\theta,\varphi} = \partial_\theta A_\varphi,
  \label{eq:EM_8}
\end{eqnarray}
where $A_\mu$= ($A_t$,$A_r$,$A_\theta$,$A_\varphi$) 
denotes the vector potential;
$F_{\mu\nu}= \partial_\mu A_\nu - \partial_\nu A_\mu$.
Note that $F_{r \theta,\varphi}+F_{\theta\varphi,r}+F_{\varphi r, \theta}=0$
(i.e., $\nabla \cdot \mbox{\boldmath$B$}=0$)
is automatically satisfied.

For an observer whose four velocity is $u^\nu$,
the electromagnetic field components are given by
\begin{equation}
  E_\mu= F_{\mu \nu} u^\nu,
\end{equation}
and
\begin{equation}
  B^\mu= \frac12 \eta_{\nu}{}^{\mu \rho \sigma} F_{\rho \sigma} u^\nu,
\end{equation}
where the completely anti-symmetric, Levi-Civita tensor density 
is defined by
\begin{equation}
   \eta_{\nu \mu \rho \sigma}
 =
  -\eta_{\mu \nu \rho \sigma}
 =-\eta_{\rho \mu \nu \sigma}
 =-\eta_{\sigma \mu \rho \nu}
 =
  -\eta_{\nu \rho \mu \sigma}
 =-\eta_{\nu \sigma \rho \mu}
 =-\eta_{\nu \mu \sigma \rho}
 \label{eq:LeviCivita_1}
\end{equation}
and
\begin{equation}
   \eta_{t r \theta \varphi} = -\frac{1}{\eta^{t r \theta \varphi}} = \sqrt{-g}.
 \label{eq:LeviCivita_2}
\end{equation}
Thus, for a distant static observer whose four velocity is
$\xi^\mu=(1,0,0,0)$, that is, the time-like Killing vector, 
the electric field components are obtained by
\begin{equation}
  E_r = F_{r t}, \quad
  E_\theta = F_{\theta t}, \quad
  E_\varphi= F_{\varphi t}
  \label{eq:E_field}
\end{equation}
in the Boyer-Lindquist coordinate.
The magnetic field becomes
\begin{eqnarray}
  B^r 
  &=& 
  \frac{-g_{tt} F_{\theta \varphi} + g_{t\varphi} F_{\theta t}}
       {\sqrt{-g}}, \quad
  \nonumber\\
  B^\theta 
  &=&
  \frac{-g_{tt} F_{\varphi r} + g_{t\varphi} F_{r t}}
       {\sqrt{-g}}, \quad
  \nonumber\\
  B_\varphi 
  &=&
  -\sqrt{-g} F^{r \theta}. \quad
  \label{eq:B_field}
\end{eqnarray}

Substituing equations~(\ref{eq:EM_4}), (\ref{eq:EM_5}),
(\ref{eq:EM_7}) and (\ref{eq:EM_8}) into equation~(\ref{eq:Poisson_2}),
we obtain the Gauss's law:
\begin{eqnarray}
  \frac{1}{\Sigma} 
  \frac{\partial}{\partial r}
  \left( \frac{A}{\Sigma} \frac{\partial A_t}{\partial r} 
  \right)
 &+&
  \frac{1}{\Delta\Sigma\sin\theta} 
  \frac{\partial}{\partial \theta}
  \left( \frac{A}{\Sigma} \sin\theta 
         \frac{\partial A_t}{\partial \theta} \right)
 +
 \frac{1}{\Sigma} 
  \frac{\partial}{\partial r}
  \left( \frac{2Mar}{\Sigma} \frac{\partial A_\varphi}{\partial r}
  \right) 
  \nonumber\\
  &+&
  \frac{1}{\Delta\Sigma\sin\theta} 
  \frac{\partial}{\partial \theta}
  \left( \frac{2Mar}{\Sigma} \sin\theta 
         \frac{\partial A_\varphi}{\partial \theta}
  \right)
  = 4\pi \rho.
 \label{eq:Poisson_3}
\end{eqnarray}
It follows that this Poisson equation contains only $A_t$
(i.e., the scalar potential) if the BH is non-rotating 
(i.e., $a=0$).
However, around a rotating BH (i.e., if $a \ne 0$), 
$A_\varphi$ comes into the Gauss's law.
We thus need one more differential equation
that contains both $A_t$ and $A_\varphi$.

\subsection[]{The Biot and Savart law}
\label{sec:BS_law}
To obtain an independent constraint on $A_t$ and $A_\varphi$,
we consider the Biot and Savart law.
Putting $\nu=\varphi$ in equation~(\ref{eq:Poisson_1}), we obtain
\begin{equation}
  \frac{1}{\sqrt{-g}} \partial_{\mu} (\sqrt{-g} F^{\varphi \mu})
  =\frac{4\pi}{c} J^\varphi,
  \label{eq:BS_1}
\end{equation}
Thus, equations~(\ref{eq:EM_1})--(\ref{eq:EM_8}) give
\begin{eqnarray}
\lefteqn{
 -\frac{1}{\sin\theta} 
  \frac{\partial}{\partial r}
  \left( \frac{\Delta-a^2\sin^2\theta}{\Sigma} 
  \frac{\partial A_\varphi}{\partial r} 
  \right)
 -\frac{1}{\Delta} 
  \frac{\partial}{\partial \theta}
  \left( \frac{\Delta-a^2\sin^2\theta}{\Sigma\sin\theta} 
         \frac{\partial A_\varphi}{\partial \theta} \right)
        }
 \nonumber\\
 &+&\frac{\partial}{\partial r}
  \left( \frac{2Mar\sin\theta}{\Sigma} 
  \frac{\partial A_t}{\partial r}
  \right) 
  +\frac{1}{\Delta} 
  \frac{\partial}{\partial \theta}
  \left( \frac{2Mar\sin\theta}{\Sigma} 
         \frac{\partial A_t}{\partial \theta}
  \right)
  = 4\pi \Sigma\sin\theta J^\varphi,
 \label{eq:BS_2}
\end{eqnarray}
where $c=1$ is used.
It follows that this equation contains only $A_\varphi$
(i.e., the azimuthal component of the vector potential) 
if $a=0$.
However, if $a \ne 0$, 
we must simultaneously solve the two second-order partial differential
equations~(\ref{eq:Poisson_3}) and (\ref{eq:BS_2})
for $A_t$ and $A_\varphi$.

It should be noted that the coefficients
of the $\partial_r{}^2 A_\varphi$ and the $\partial_\theta{}^2 A_\varphi$
terms in equation~(\ref{eq:BS_2})
change sign at the static limit, where $g_{tt}$ vanishes.
Since the coefficients of these two highest-order derivative terms 
change sign with respect to those of the other 
two highest-order derivative terms 
(i.e., the $\partial_r{}^2 A_t$ and  the $\partial_\theta{}^2 A_t$ terms),
the solution diverges during iterations
from the inside of the static limit 
(i.e., within the so-called \lq\lq ergosphere'')
if we impose boundary conditions in the same way as
standard elliptic type partial differential equations.
This ill behaviour is incurred 
because a static observer (with respect to the star) 
becomes unphysical within the ergosphere.

There are many ways to overcome this ill behaviour.
In the present paper, we solve this issue by adopting
the ZAMO as a physical observer.
Using the ZAMO's tetrad,
equations~(\ref{eq:rotationg_e0z})--
(\ref{eq:rotationg_e3z}),
we can replace $A_t$ with the ZAMO-measured
scalar potential, $A_{\hat{t}}$
(eq.~[\ref{eq:At}]),
and obtain the two elliptic-type 
equations~(\ref{eq:BS_3}) and (\ref{eq:Poisson_4}),
which describe the initial electromagnetic fields
at $t=0$.

\section{Pair production rate}
\label{sec:app_ADAF}
Using the self-similar analytical solution of Newtonian ADAF model
\citep{Mahadevan:1997:ApJ},
we find that the photons are mitted at the rate 
\begin{equation}
  q_{\rm ff}
  = 1.8 \times 10^2 \theta_{\rm e}
    \dot{m}^2 M_9{}^{-2} \left(\frac{r}{2M}\right)^{-3}
    \mbox{ergs s}^{-1} \mbox{cm}^{-3}
\end{equation}
by Bremsstrahlung,
where $\theta_{\rm e} \equiv kT_{\rm e}/m_{\rm e}c^2$
denotes the dimensionless electron temperature.
The luminosity of this Bremsstrahlung emission can be computed as
\begin{equation}
  L_{\rm ff}(r)
  = 2\pi \int_{\theta_1}^{\theta_2} 
    \int_{r_{\rm min}}^r q_{\rm ff}(r) r^2 \sin\theta dr d\theta
  \approx
    2\pi \times 1.8 \times 10^2 \theta_{\rm e}
    \dot{m}^2 M_9{}^{-2} (2M)^3 \ln \left(\frac{r}{2M}\right),
\end{equation}
where $\theta_1 \sim 60^\circ$ and $\theta_2 \sim 120^\circ$
denote the upper and lower boundary colatitudes of the ADAF.

In general, if one photon species 
(e.g., ADAF Bremsstrahlung photons or gap-emitted gamma-rays)
collide with
another photon species
(e.g., ADAF Bremsstrahlung photons or ADAF synchrotron photons),
the pair proction rate is given by
\begin{equation}
  \dot{n}_\pm(r)
  = \int d\nu_\gamma \alpha_{\gamma\gamma}
    \frac{1}{c} \int \frac{I_\nu}{h\nu} d\Omega_\gamma,
  \label{eq:Ngamma}
\end{equation}
where $d\Omega_\gamma$ refers to the photon propagation solid angle,
and
\begin{equation}
  \alpha_{\gamma\gamma}
  = (1-\mu) 
    \int_{\epsilon_{\rm th}}^\infty
      \frac{dF_\gamma}{d\epsilon_\gamma} \sigma_{\gamma\gamma}
      d \epsilon_\gamma
  \approx 
     (1-\mu) F_\gamma \sigma_{\gamma\gamma}
  \label{eq:absorption_coeff}
\end{equation}
denotes the photon-photon absorption coefficient,
and 
\begin{equation}
  F_\gamma 
  \equiv \int d\nu_\gamma \int \frac{I_\nu}{h\nu} d\Omega_\gamma
\end{equation}
does the photon number flux;
$I_\nu$, $\nu_\gamma$ and $\mu$ show
the specific intensity, photon frequency, 
and the cosine of the photon collision angles, respectively.

In the present case, the two species are the same ADAF
Bremsstrahlung photons.
Thus, we obtain
\begin{equation}
  \dot{n}_\pm(r)
  \approx (1-\mu) \sigma_{\gamma\gamma}
    \frac{F_\gamma{}^2}{c}
  \label{eq:Ngamma3}
\end{equation}
Evaluating the flux with the luminosity by
\begin{equation}
  F_\gamma(r) \approx \frac{L_{\rm ff}(r)}{2\pi r^2 \epsilon_\gamma},
\end{equation}
where the photon energy can be estimated to be
$\epsilon_\gamma \approx 3 \theta_{\rm e} m_{\rm e} c^2$,
and assuming that the pair production rate is uniform within $r<4M$,
we obtain equation~(\ref{eq:dotN2}),
where the relative velocity factor is evaluated as
$1-\mu= 0.2$,
and the pair-production cross section as
the $20\%$ of the Thomson cross section,
$\sigma_{\gamma\gamma}=0.2 \sigma_{\rm T}$.

\section{Covariance of the radiation-reaction force}
\label{sec:app_covaniance}

We explain here that the radiation reaction force contains
not only those obtained in a flat space
\citep{cerutti12,cerutti13},
\begin{equation}
    P_{\rm rad, flat}
    = \frac23 e^2 c \frac{q^2}{m^2}
      \gamma^2 
    \left[ -\left( \frac{\mbox{\boldmath$v$}}{c}
                            \cdot \tilde{\mbox{\boldmath$E$}}
                     \right)^2
                   +\left( \tilde{\mbox{\boldmath$E$}}
                          + \frac{\mbox{\boldmath$v$}}{c}
                            \times \tilde{\mbox{\boldmath$B$}}
                    \right)^2
             \right],
  \qquad
  \label{eq:P_flat}
\end{equation}
in locally homogeneous $\tilde{\mbox{\boldmath$E$}}$
and $\tilde{\mbox{\boldmath$B$}}$ fields,
but also contains the power of radiative processes that result
from any kinds of acceleration.
To find this, we write down equations~(\ref{eq:EOM})
in a covariat form,
\begin{equation}
  \frac{d u^\mu}{d\lambda}
  = -\Gamma^\mu{}_{\rho\sigma} u^\rho u^\sigma
    +\frac{q}{m} F^\mu{}_\rho u^\rho,
\end{equation}
where the friction term is dropped here in the right-hand side
to draw the conclusion from the first principles.
Both the inertial forces (e.g., the centrifugal and the Colioris forces) 
and the gravitational forces (due to the spacetime curvature)
are included in the connection coefficients, $\Gamma^\mu{}_{\rho\sigma}$.
For example, when a charge (or its guiding center)
moves along a curved magnetic field line in a flat space,
we can introduce a local cylindrical (or polar) coordinate system
whose origin resides at the center of the curved path of 3D particle motion.
In this coordinate, 
the $\Gamma^\mu{}_{\rho\sigma} u^\rho u^\sigma$ term gives
the centripetal acceleration, $a_\perp= c^2/R_{\rm c}$, where
$R_{\rm c}$ refers to the curvature radius of the 3D particle motion.
Since equation~(\ref{eq:def_Prad}) is covariant, the effect
of such curvature radiation is included in any frame of refence.
Accordingly, we can take account of 
radiation effects due to any acceleration.
Note that a charged particle does radiate
by gravitational acceleration 
irrespective of the frame of reference.

In actual program, we can compute the radiation-reaction force
$F_{\rm rad}^j$ iteratively.
Namely, \\
(1) we first set $F_{\rm rad}^j=0$ and update
    $du^j/d\lambda$ and $du^0/d\lambda=\gamma$
    by equation~(\ref{eq:EOM}),\\
(2) second, we subsitiute the solved $u^\mu$ 
    in equation~(\ref{eq:def_Prad}) to update $F_{\rm rad}^j$,\\
(3) third, use the updated $F_{\rm rad}^j$ and solve 
    equation~(\ref{eq:EOM}) for $u^\mu$ again,\\
(4) iterate steps (2) and (3) until $F_{\rm rad}^j$ saturates.
If we use the $F_{\rm rad}^j$ obtained
in the previous step for each particle and start from
step (3), the number of iterations can be reduced.

\section[]{Discretization of basic equations}
\label{sec:discretization}
We discretize the Maxwell 
equations~(\ref{eq:Maxwell_1b})--(\ref{eq:Maxwell_3b})
adopting the Yee lattice.
Electric field components, $E$ and $D$,
are evaluated at half integer time steps,
e.g., at $t \pm (1/2)\Delta_t$, 
and the magnetic field compoent, $B$,
is evaluated at integer time step 
at time $t+\Delta_t$.
Accordingly, equations~(\ref{eq:Maxwell_1b})--(\ref{eq:Maxwell_3b})
are discretized as follows:
\begin{eqnarray}
  \lefteqn{
  B^{n+1}{}_{i+1/2,j+1/2} -B^{n}{}_{i+1/2,j+1/2} }
  \nonumber\\
  = 
    && -c_1 \nu_x \left( E^{n+1/2}{}_{i+1,j+1/2}-E^{n+1/2}{}_{i,j+1/2} \right)
    +c_2 \nu_y \left( D^{n+1/2}{}_{i+1/2,j+1}-D^{n+1/2}{}_{i+1/2,j} \right) ,
  \label{eq:Maxwell_1c}
\end{eqnarray}
\begin{eqnarray}
  \lefteqn{
  D^{n+1/2}{}_{i+1/2,j} -D^{n-1/2}{}_{i+1/2,j} }
  \nonumber\\
  =
    && c_3 \nu_y \left( B^{n}{}_{i+1/2,j+1/2}-B^{n}{}_{i+1/2,j-1/2} \right)
    -4\pi (\Delta_t) \left. \frac{\Sigma^2}{A}
                     \right|_{i+1/2,j}
     (J^r)^{n}_{i+1/2,j},
  \label{eq:Maxwell_2c}
\end{eqnarray}
\begin{eqnarray}
  \lefteqn{
  E^{n+1/2}{}_{i,j+1/2} -E^{n-1/2}{}_{i,j+1/2} }
  \nonumber\\
  = 
    && -c_4 \nu_x \left( B^{n}{}_{i+1/2,j+1/2}-B^{n}{}_{i-1/2,j+1/2} \right)
    -4\pi (\Delta_t) \left. \frac{\Sigma^2\Delta\sin\theta}{A}
                     \right|_{i+1/2,j}
     (J^\theta)^{n}_{i,j+1/2},
  \label{eq:Maxwell_3c}
\end{eqnarray}
where the super- and sub-scripts denote
the temporal and spatial labels, respectively;
and the Courant numbers are defined by
\begin{equation}
  \nu_x \equiv \frac{\Delta_t}{\Delta_x},
  \label{eq:Courant_x}
\end{equation}
\begin{equation}
  \nu_y \equiv \frac{\Delta_t}{\Delta_y}.
  \label{eq:Courant_y}
\end{equation}
Here, $\Delta_t$ denotes the time step, 
$\Delta_x$ denotes the constant interval of the radial tortoise coordinate, 
and $\Delta_y = \sin\theta \Delta_\theta$
denotes the constant azimuthal grid interval.
Denoting the inner and outer boundary positions as $r_\ast{}^{\rm min}$
and $r_\ast{}^{\rm max}$, we obtain 
$\Delta_x= (r_\ast{}^{\rm max}-r_\ast{}^{\rm min})/(N_x-1)$,
where $N_x$ denotes the number of radial grids.
Since we consider the colatitudes range 
$0 < 1-\cos\theta < 2$, 
we obtain $\Delta_y= 2/(N_y-1)$, where
$N_y$ denotes the number of meridional grids.

As for the Hamilton-Jacobi equations~(\ref{eq:EOM}) and 
(\ref{eq:drdt})--(\ref{eq:dphdt}), 
we evaluate the particle position ($r$,$\theta$,$\varphi$) 
at half integer time steps, and 
the momentum ($u^r$,$u^\theta$,$u^\varphi$, and $\gamma$) 
at integer time steps.
Namely, particle's momentum evolves by (eq.~[\ref{eq:EOM}])
\begin{equation}
  \mbox{\boldmath$u$}^{n+1}-\mbox{\boldmath$u$}^{n}
  = (\Delta_t) \mbox{\boldmath$F$}(t^{n+1/2},\mbox{\boldmath$x$}^{n+1/2},
                                           \mbox{\boldmath$u$}^{n+1/2}),
  \label{eq:EOM_1}
\end{equation}
where $F_i$ represents the right-hand side of equation~(\ref{eq:EOM}).

On the other hand, particle's position evolves by 
(eq.~[\ref{eq:drdt}]--[\ref{eq:dphdt}])
\begin{equation}
  r^{n+1/2}= r^{n-1/2}+ \left( \frac{u^r}{u^t} \right)^{n},
\end{equation}
\begin{equation}
  \theta^{n+1/2}= \theta^{n-1/2}
                + \left( \frac{u^\theta}{u^t} \right)^{n},
\end{equation}
\begin{equation}
  \varphi^{n+1/2}= \varphi^{n-1/2}
                + \left( \frac{u^\varphi}{u^t} \right)^{n}
\end{equation}
The Boyer-Lindquist radial coordinate $r$ 
can be readily converted into $x=r_\ast$ by equation~(\ref{eq:tortoise}),
whereas $\theta$ is related to $y$ by $y=1-\cos\theta$.

To find the initial, stationary solution,
we divide the tortoise coordinate $x$ uniformly
between $-15.8180 < x < 20.0000$,
and the meridional coordinate $y$ uniformly 
between $0 < y=1-\cos\theta < 2$.
Accordingly, the Boyer-Lindquist coordinates are divided non-uniformly as
$r_1=1.4651M$, $r_2=1.4656M$, $r_3=1.4662M$, $\ldots$,
$r_{599}=19.9462M$, $r_{600}=20.0000M$, and
$\theta_1=0^\circ$, $\theta_2=2.6160^\circ$, 
$\theta_3=3.7000^\circ$, $\theta=4.5319^\circ$,
$\ldots$,
$\theta_{959}= 89.9104^\circ$,
$\theta_{960}= 89.9701^\circ$,
$\theta_{961}= 90.0298^\circ$,
$\ldots$,
$\theta_{1918}=(180.0000-3.7000)^\circ$,
$\theta_{1919}=(180.0000-2.6160)^\circ$
$\theta_{1920}=180.0000^\circ$.

To perform PIC simulations,
we must restrict the radial range small compared to the stationary case,
because the Gauss's and Biot-Savart laws are solved
for the toroidal current distribution within $r<20M$,
which makes the solved poloidal electromagnetic fields
be accurate typically only within $r < 14M$.
The actual value of the right-hand side 
(i.e., $14M$ in this case) depends on 
the $r$ dependence of $J^\varphi{}_{\rm eq}$.
For a split-monople case, $J^\varphi{}_{\rm eq} \propto r^{-4}$, 
or a paraboloidal case, $J^\varphi{}_{\rm eq} \propto r^{-3}$,
the outer radial boundary of $r \sim 14M$ is conservatively justified.
For PIC simulations, 
we thus adopt the inner 960 points 
to restrict the simulation range within
$x_1=-15.8180 < x < x_{960}= 12.8244$, 
or equivalently,
$r_1=1.4646M < r < r_{960}=13.6854M$.
For meridional coordinate $y$, and adopt the same gridding
as in the stationary case.
We checked that the results little change when we halve the resolution
either in radial or meridional direction, or when we halve 
the creation rate $1/k_{\rm create}$
(and hence the number of particles per cell).

\section[]{Current deposit}
\label{sec:current}
In our present 2D PIC simulations,
we adopt a grid consisting of rectangles of unit size,
and place on it unit rectangular charges,
where each unit rectangle grid cell has 
an area $\Delta_x \Delta_y$
(Appendix~\ref{sec:discretization}).
The total amount of charges is assumed to be the uniformly
distributed over its 2D surface.
As a charge moves, each grid cell boundary sweeps a fraction
of the particle's unit area surface,
giving the electric current flowing across the boundary.
Using this \lq area weighting' \citep{villa92},
we sum up the currents carried by individual particles 
on each grid boundary.

We should notice here that we take account of the
particle motion only in the poloidal plane
when we compute the current density,
although the particles' EOM is solved 
three-dimensionally.
Thus, we cannot compute the toroidal current density,
$J^\varphi$, in our area weighting method.
It follows that the time evolution 
of $E_\varphi$ cannot be solved from the Ampere's law.
Accordingly, the evolution of $B^r$ or $B^\theta$ 
cannot be solved either,
because $r$ and $\theta$ components of the Faraday law
contain $\partial_\theta E_\varphi$ and 
$\partial_r E_\varphi$, respectively.
Thus, we are able to solve only the three components
(i.e., $B$, $D$, and $E$)
of the electromagnetic field by
equations~(\ref{eq:Maxwell_1b})--(\ref{eq:Maxwell_3b})
in the present 2D scheme, 
unless $J^\varphi$ is constructed from 
particle's toroidal motion
by a \lq volume weighting' method,
which is out of the scope of the present paper.

\bibliographystyle{aasjournal}



\end{document}